\input harvmac.tex
\input epsf.tex
\input amssym
\input ulem.sty
\input graphicx.tex
%\draftmode

\let\includefigures=\iftrue
\let\useblackboard=\iftrue
\newfam\black

\def\figin{\epsfcheck\figin}\def\figins{\epsfcheck\figins}
\def\epsfcheck{\ifx\epsfbox\UnDeFiNeD
\message{(NO epsf.tex, FIGURES WILL BE IGNORED)}
\gdef\figin##1{\vskip2in}\gdef\figins##1{\hskip.5in}% blank space instead
\else\message{(FIGURES WILL BE INCLUDED)}%
\gdef\figin##1{##1}\gdef\figins##1{##1}\fi}
\def\DefWarn#1{}
\def\figinsert{\goodbreak\midinsert}
\def\ifig#1#2#3{\DefWarn#1\xdef#1{fig.~\the\figno}
\writedef{#1\leftbracket fig.\noexpand~\the\figno} %
\figinsert\figin{\centerline{#3}}\medskip\centerline{\vbox{\baselineskip12pt
\advance\hsize by -1truein\noindent\footnotefont{\bf
Fig.~\the\figno:} #2}}
\bigskip\endinsert\global\advance\figno by1}

% TO INCLUDE FIGURES DO AS BELOW
%\ifig\LABEL{  WRITE CAPTION } {\epsfxsize1.5in\epsfbox{FILENAME.eps}}

%Figure Stuff
\includefigures
\message{If you do not have epsf.tex (to include figures),}
\message{change the option at the top of the tex file.}
\input epsf
\def\figin{\epsfcheck\figin}\def\figins{\epsfcheck\figins}
\def\epsfcheck{\ifx\epsfbox\UnDeFiNeD
\message{(NO epsf.tex, FIGURES WILL BE IGNORED)}
\gdef\figin##1{\vskip2in}\gdef\figins##1{\hskip.5in}% blank space instead
\else\message{(FIGURES WILL BE INCLUDED)}%
\gdef\figin##1{##1}\gdef\figins##1{##1}\fi}
\def\DefWarn#1{}
\def\figinsert{\goodbreak\midinsert}
\def\ifig#1#2#3{\DefWarn#1\xdef#1{fig.~\the\figno}
\writedef{#1\leftbracket fig.\noexpand~\the\figno}%
\figinsert\figin{\centerline{#3}}\medskip\centerline{\vbox{
\baselineskip12pt\advance\hsize by -1truein
\noindent\footnotefont{\bf Fig.~\the\figno:} #2}}
%\bigskip
\endinsert\global\advance\figno by1}
%%%
\else
\def\ifig#1#2#3{\xdef#1{fig.~\the\figno}
\writedef{#1\leftbracket fig.\noexpand~\the\figno}%
%\figinsert\figin{\centerline{#3}}\medskip
%\centerline{\vbox{\baselineskip12pt
%\advance\hsize by -1truein\noindent
%\footnotefont{\bf Fig.~\the\figno:} #2}}
%\bigskip\endinsert
\global\advance\figno by1} \fi

\def\figin{\epsfcheck\figin}\def\figins{\epsfcheck\figins}
\def\epsfcheck{\ifx\epsfbox\UnDeFiNeD
\message{(NO epsf.tex, FIGURES WILL BE IGNORED)}
\gdef\figin##1{\vskip2in}\gdef\figins##1{\hskip.5in}% blank space instead
\else\message{(FIGURES WILL BE INCLUDED)}%
\gdef\figin##1{##1}\gdef\figins##1{##1}\fi}
\def\DefWarn#1{}
\def\figinsert{\goodbreak\midinsert}
\def\ifig#1#2#3{\DefWarn#1\xdef#1{fig.~\the\figno}
\writedef{#1\leftbracket fig.\noexpand~\the\figno} %
\figinsert\figin{\centerline{#3}}\medskip\centerline{\vbox{\baselineskip12pt
\advance\hsize by -1truein\noindent\footnotefont{\bf
Fig.~\the\figno:} #2}}
\bigskip\endinsert\global\advance\figno by1}

\def \pa {\partial}
\def \pd {\partial}

\def\OO{{\cal OO}}

\catcode`\@=11
\def\slash#1{\mathord{\mathpalette\c@ncel{#1}}}
\overfullrule=0pt

\def\BB{{\cal B}}

\def\GG{{\cal G}}

\def\OO{{\cal O}}

\def\underrel#1\over#2{\mathrel{\mathop{\kern\z@#1}\limits_{#2}}}

\catcode`\@=12

%%%%%%%%%%%%%%%%%%%%%%%%%%%%%%%%%%%%%%%%%%%%%%%%%%%%%%%%%%%%%%

\def \cosh{{\rm cosh}}

%%%%%%%%%%%%%%%%%%%%%%%%%%%%%%%%%%%%%%%%%%%%%%%%%%%%%%%%%%%%%%
% new defs:

\def\zbar{{\bar z}}

\def\lra{\leftrightarrow}

\def\hbar{{\bar h}}

\def\DL{{\Delta_{L}}}
\def\DH{{\Delta_{H}}}

\def\OL{{\OO_{L}}}
\def\OH{{\OO_{H}}}

\def\ebar{{\bar e}}

\def\lzbarone{{\underrel{\approx}\over{\bar{z}\rightarrow 1}}}

\def\rhoequal{{\underrel{\approx}\over{\rho\to\infty}}}

%\KulaxiziDXO
\lref\KulaxiziDXO{
  M.~Kulaxizi, G.~S.~Ng and A.~Parnachev,
  ``Black Holes, Heavy States, Phase Shift and Anomalous Dimensions,''
SciPost Phys.\  {\bf 6}, 065 (2019).
[arXiv:1812.03120 [hep-th]].
%%CITATION = arXiv:1812.03120%%
}

%\KulaxiziTKD
\lref\KulaxiziTKD{
  M.~Kulaxizi, G.~S.~Ng and A.~Parnachev,
  ``Subleading Eikonal, AdS/CFT and Double Stress Tensors,''
[arXiv:1907.00867 [hep-th]].
%%CITATION = arXiv:1907.00867%%
}

%\KarlssonQFI
\lref\KarlssonQFI{
  R.~Karlsson, M.~Kulaxizi, A.~Parnachev and P.~Tadi´c,
  ``Black Holes and Conformal Regge Bootstrap,''
[arXiv:1904.00060 [hep-th]].
%%CITATION = arXiv:1904.00060%%
}
%\KarlssonDBD
\lref\KarlssonDBD{
  R.~Karlsson, M.~Kulaxizi, A.~Parnachev and P.~Tadi´c,
  ``Leading Multi-Stress Tensors and Conformal Bootstrap,''
[arXiv:1909.05775 [hep-th]].
%%CITATION = arXiv:1909.05775%%
}
%\FitzpatrickEFK
\lref\FitzpatrickEFK{
  A.~L.~Fitzpatrick, K.~W.~Huang and D.~Li,
  ``Probing Universalities in d>2 CFTs: from Black Holes to Shockwaves,''
[arXiv:1907.10810 [hep-th]].
%%CITATION = arXiv:1907.10810%%
}
%\PenedonesNS
\lref\PenedonesNS{
  J.~Penedones,
  ``High Energy Scattering in the AdS/CFT Correspondence,''
[arXiv:0712.0802 [hep-th]].
%%CITATION = arXiv:0712.0802%%
}
%\KravchukHTV
\lref\KravchukHTV{
  P.~Kravchuk and D.~Simmons-Duffin,
  ``Light-ray operators in conformal field theory,''
JHEP {\bf 1811}, 102 (2018).
[arXiv:1805.00098 [hep-th]].
%%CITATION = arXiv:1805.00098%%
}
%\KulaxiziIXA
\lref\KulaxiziIXA{
  M.~Kulaxizi, A.~Parnachev and A.~Zhiboedov,
  ``Bulk Phase Shift, CFT Regge Limit and Einstein Gravity,''
JHEP {\bf 1806}, 121 (2018).
[arXiv:1705.02934 [hep-th]].
%%CITATION = arXiv:1705.02934%%
}
%\DolanHV
\lref\DolanHV{
  F.~A.~Dolan and H.~Osborn,
  ``Conformal partial waves and the operator product expansion,''
Nucl.\ Phys.\ B {\bf 678}, 491 (2004).
[hep-th/0309180].
%%CITATION = DAMTP-03-91%%
}
%\DolanUT
\lref\DolanUT{
  F.~A.~Dolan and H.~Osborn,
  ``Conformal four point functions and the operator product expansion,''
Nucl.\ Phys.\ B {\bf 599}, 459 (2001).
[hep-th/0011040].
%%CITATION = hep-th/0011040%%
}
%\FitzpatrickZQZ
\lref\FitzpatrickZQZ{
  A.~L.~Fitzpatrick and K.~W.~Huang,
  ``Universal Lowest-Twist in CFTs from Holography,''
JHEP {\bf 1908}, 138 (2019).
[arXiv:1903.05306 [hep-th]].
%%CITATION = arXiv:1903.05306%%
}
%\HuangFOG
\lref\HuangFOG{
  K.~W.~Huang,
  ``Stress-tensor commutators in conformal field theories near the lightcone,''
Phys.\ Rev.\ D {\bf 100}, no. 6, 061701 (2019).
[arXiv:1907.00599 [hep-th]].
%%CITATION = arXiv:1907.00599%%
}
%\RattazziPE
\lref\RattazziPE{
  R.~Rattazzi, V.~S.~Rychkov, E.~Tonni and A.~Vichi,
  ``Bounding scalar operator dimensions in 4D CFT,''
JHEP {\bf 0812}, 031 (2008).
[arXiv:0807.0004 [hep-th]].
%%CITATION = arXiv:0807.0004%%
}
%\MaldacenaRE
\lref\MaldacenaRE{
  J.~M.~Maldacena,
  ``The Large N limit of superconformal field theories and supergravity,''
Int.\ J.\ Theor.\ Phys.\  {\bf 38}, 1113 (1999), [Adv.\ Theor.\ Math.\ Phys.\  {\bf 2}, 231 (1998)].
[hep-th/9711200].
%%CITATION = HUTP-97-A097%%
}
%\WittenQJ
\lref\WittenQJ{
  E.~Witten,
  ``Anti-de Sitter space and holography,''
Adv.\ Theor.\ Math.\ Phys.\  {\bf 2}, 253 (1998).
[hep-th/9802150].
%%CITATION = hep-th/9802150%%
}
%\GubserBC
\lref\GubserBC{
  S.~S.~Gubser, I.~R.~Klebanov and A.~M.~Polyakov,
  ``Gauge theory correlators from noncritical string theory,''
Phys.\ Lett.\ B {\bf 428}, 105 (1998).
[hep-th/9802109].
%%CITATION = hep-th/9802109%%
}
%\CornalbaXM
\lref\CornalbaXM{
  L.~Cornalba, M.~S.~Costa, J.~Penedones and R.~Schiappa,
  ``Eikonal Approximation in AdS/CFT: Conformal Partial Waves and Finite N Four-Point Functions,''
Nucl.\ Phys.\ B {\bf 767}, 327 (2007).
[hep-th/0611123].
%%CITATION = hep-th/0611123%%
}
%\KomargodskiEK
\lref\KomargodskiEK{
  Z.~Komargodski and A.~Zhiboedov,
  ``Convexity and Liberation at Large Spin,''
JHEP {\bf 1311}, 140 (2013).
[arXiv:1212.4103 [hep-th]].
%%CITATION = arXiv:1212.4103%%
}
%\FitzpatrickYX
\lref\FitzpatrickYX{
  A.~L.~Fitzpatrick, J.~Kaplan, D.~Poland and D.~Simmons-Duffin,
  ``The Analytic Bootstrap and AdS Superhorizon Locality,''
JHEP {\bf 1312}, 004 (2013).
[arXiv:1212.3616 [hep-th]].
%%CITATION = arXiv:1212.3616%%
}
%\CornalbaXK
\lref\CornalbaXK{
  L.~Cornalba, M.~S.~Costa, J.~Penedones and R.~Schiappa,
  ``Eikonal Approximation in AdS/CFT: From Shock Waves to Four-Point Functions,''
JHEP {\bf 0708}, 019 (2007).
[hep-th/0611122].
%%CITATION = hep-th/0611122%%
}
%\CornalbaXM
\lref\CornalbaXM{
  L.~Cornalba, M.~S.~Costa, J.~Penedones and R.~Schiappa,
  ``Eikonal Approximation in AdS/CFT: Conformal Partial Waves and Finite N Four-Point Functions,''
Nucl.\ Phys.\ B {\bf 767}, 327 (2007).
[hep-th/0611123].
%%CITATION = hep-th/0611123%%
}
%\CornalbaZB
\lref\CornalbaZB{
  L.~Cornalba, M.~S.~Costa and J.~Penedones,
  ``Eikonal approximation in AdS/CFT: Resumming the gravitational loop expansion,''
JHEP {\bf 0709}, 037 (2007).
[arXiv:0707.0120 [hep-th]].
%%CITATION = arXiv:0707.0120%%
}
%\CornalbaAX
\lref\CornalbaAX{
  L.~Cornalba, M.~S.~Costa and J.~Penedones,
  ``Deep Inelastic Scattering in Conformal QCD,''
JHEP {\bf 1003}, 133 (2010).
[arXiv:0911.0043 [hep-th]].
%%CITATION = arXiv:0911.0043%%
}
%\CostaCB
\lref\CostaCB{
  M.~S.~Costa, V.~Goncalves and J.~Penedones,
  ``Conformal Regge theory,''
JHEP {\bf 1212}, 091 (2012).
[arXiv:1209.4355 [hep-th]].
%%CITATION = arXiv:1209.4355%%
}
%\FitzpatrickVUA
\lref\FitzpatrickVUA{
  A.~L.~Fitzpatrick, J.~Kaplan and M.~T.~Walters,
  ``Universality of Long-Distance AdS Physics from the CFT Bootstrap,''
JHEP {\bf 1408}, 145 (2014).
[arXiv:1403.6829 [hep-th]].
%%CITATION = arXiv:1403.6829%%
}
%\FitzpatrickZHA
\lref\FitzpatrickZHA{
  A.~L.~Fitzpatrick, J.~Kaplan and M.~T.~Walters,
  ``Virasoro Conformal Blocks and Thermality from Classical Background Fields,''
JHEP {\bf 1511}, 200 (2015).
[arXiv:1501.05315 [hep-th]].
%%CITATION = arXiv:1501.05315%%
}
%\HijanoRLA
\lref\HijanoRLA{
  E.~Hijano, P.~Kraus and R.~Snively,
  ``Worldline approach to semi-classical conformal blocks,''
JHEP {\bf 1507}, 131 (2015).
[arXiv:1501.02260 [hep-th]].
%%CITATION = arXiv:1501.02260%%
}
%\HijanoQJA
\lref\HijanoQJA{
  E.~Hijano, P.~Kraus, E.~Perlmutter and R.~Snively,
  ``Semiclassical Virasoro blocks from AdS$_{3}$ gravity,''
JHEP {\bf 1512}, 077 (2015).
[arXiv:1508.04987 [hep-th]].
%%CITATION = arXiv:1508.04987%%
}
%\FitzpatrickFOA
\lref\FitzpatrickFOA{
  A.~L.~Fitzpatrick, J.~Kaplan, M.~T.~Walters and J.~Wang,
  ``Hawking from Catalan,''
JHEP {\bf 1605}, 069 (2016).
[arXiv:1510.00014 [hep-th]].
%%CITATION = arXiv:1510.00014%%
}
%\CollierEXN
\lref\CollierEXN{
  S.~Collier, Y.~Gobeil, H.~Maxfield and E.~Perlmutter,
  ``Quantum Regge Trajectories and the Virasoro Analytic Bootstrap,''
JHEP {\bf 1905}, 212 (2019).
[arXiv:1811.05710 [hep-th]].
%%CITATION = arXiv:1811.05710%%
}
%\BeskenJYW
\lref\BeskenJYW{
  M.~Besken, S.~Datta and P.~Kraus,
  ``Semi-classical Virasoro blocks: proof of exponentiation,''
[arXiv:1910.04169 [hep-th]].
%%CITATION = arXiv:1910.04169%%
}
%\CamanhoAPA
\lref\CamanhoAPA{
  X.~O.~Camanho, J.~D.~Edelstein, J.~Maldacena and A.~Zhiboedov,
  ``Causality Constraints on Corrections to the Graviton Three-Point Coupling,''
JHEP {\bf 1602}, 020 (2016).
[arXiv:1407.5597 [hep-th]].
%%CITATION = arXiv:1407.5597%%
}
%\CostaTWZ
\lref\CostaTWZ{
  M.~S.~Costa, T.~Hansen and J.~Penedones,
  ``Bounds for OPE coefficients on the Regge trajectory,''
JHEP {\bf 1710}, 197 (2017).
[arXiv:1707.07689 [hep-th]].
%%CITATION = arXiv:1707.07689%%
}
%\LiLMH
\lref\LiLMH{
  D.~Li, D.~Meltzer and D.~Poland,
  ``Conformal Bootstrap in the Regge Limit,''
JHEP {\bf 1712}, 013 (2017).
[arXiv:1705.03453 [hep-th]].
%%CITATION = arXiv:1705.03453%%
}
%\PolyakovGS
\lref\PolyakovGS{
  A.~M.~Polyakov,
  ``Nonhamiltonian approach to conformal quantum field theory,''
Zh.\ Eksp.\ Teor.\ Fiz.\  {\bf 66}, 23 (1974), [Sov.\ Phys.\ JETP {\bf 39}, 9 (1974)]..
}
%\FerraraYT
\lref\FerraraYT{
  S.~Ferrara, A.~F.~Grillo and R.~Gatto,
  ``Tensor representations of conformal algebra and conformally covariant operator product expansion,''
Annals Phys.\  {\bf 76}, 161 (1973)..
}
%\FitzpatrickDLT
\lref\FitzpatrickDLT{
  A.~L.~Fitzpatrick and J.~Kaplan,
  ``Conformal Blocks Beyond the Semi-Classical Limit,''
JHEP {\bf 1605}, 075 (2016).
[arXiv:1512.03052 [hep-th]].
%%CITATION = arXiv:1512.03052%%
}
%\FitzpatrickIVE
\lref\FitzpatrickIVE{
  A.~L.~Fitzpatrick, J.~Kaplan, D.~Li and J.~Wang,
  ``On information loss in AdS$_{3}$/CFT$_{2}$,''
JHEP {\bf 1605}, 109 (2016).
[arXiv:1603.08925 [hep-th]].
%%CITATION = arXiv:1603.08925%%
}
%\AnousKSS
\lref\AnousKSS{
  T.~Anous, T.~Hartman, A.~Rovai and J.~Sonner,
  ``Black Hole Collapse in the 1/c Expansion,''
JHEP {\bf 1607}, 123 (2016).
[arXiv:1603.04856 [hep-th]].
%%CITATION = arXiv:1603.04856%%
}
%\FitzpatrickMJQ
\lref\FitzpatrickMJQ{
  A.~L.~Fitzpatrick and J.~Kaplan,
  ``On the Late-Time Behavior of Virasoro Blocks and a Classification of Semiclassical Saddles,''
JHEP {\bf 1704}, 072 (2017).
[arXiv:1609.07153 [hep-th]].
%%CITATION = arXiv:1609.07153%%
}

%\ChenNVBIP
\lref\ChenNVBIP{
  H.~Chen, C.~Hussong, J.~Kaplan and D.~Li,
  ``A Numerical Approach to Virasoro Blocks and the Information Paradox,''
[arXiv:1703.09727 [hep-th]].
%%CITATION = arXiv:1907.10810%%
}
%\FaulknerHLL
\lref\FaulknerHLL{
  T.~Faulkner and H.~Wang,
  ``Probing beyond ETH at large $c$,''
JHEP {\bf 1806}, 123 (2018).
[arXiv:1712.03464 [hep-th]].
%%CITATION = arXiv:1712.03464%%
}
%\HartmanMIA
\lref\HartmanMIA{
  T.~Hartman,
  ``Entanglement Entropy at Large Central Charge,''
[arXiv:1303.6955 [hep-th]].
%%CITATION = arXiv:1303.6955%%
}
%\AsplundCOA
\lref\AsplundCOA{
  C.~T.~Asplund, A.~Bernamonti, F.~Galli and T.~Hartman,
  ``Holographic Entanglement Entropy from 2d CFT: Heavy States and Local Quenches,''
JHEP {\bf 1502}, 171 (2015).
[arXiv:1410.1392 [hep-th]].
%%CITATION = arXiv:1410.1392%%
}
%\RobertsIFA
\lref\RobertsIFA{
  D.~A.~Roberts and D.~Stanford,
  ``Two-dimensional conformal field theory and the butterfly effect,''
Phys.\ Rev.\ Lett.\  {\bf 115}, no. 13, 131603 (2015).
[arXiv:1412.5123 [hep-th]].
%%CITATION = MIT-CTP-4626%%
}
%\CaputaQE
\lref\CaputaQE{
  P.~Caputa, J.~Simon, A.~Stikonas and T.~Takayanagi, 
  ``Quantum Entanglement of Localized Excited States at Finite Temperature,''
[arXiv:1410.2287 [hep-th]].
%%CITATION = arXiv:1907.10810%%
}
%\ChenEE
\lref\ChenEE{
  B.~Chen and J.~Wu,
  ``Holographic Entanglement Entropy For a Large Class of States in 2D CFT,''
  [arXiv:1605.06753 [hep-th]].
%%CITATION = arXiv:1907.10810%%
}
%\ChenESC
\lref\ChenESC{
  B.~Chen, J.~Wu and J.~Zhang,
  ``Holographic Description of 2D Conformal Block in Semi-classical Limit,''
  [arXiv:1609.00801 [hep-th]].
%%CITATION = arXiv:1907.10810%%
}
%\AnousBHC
\lref\AnousBHC{
  T.~Anous, T.~Hartman, A.~Rovai and J.~Sonner, 
  ``Black Hole Collapse in the 1/c Expansion,''
[arXiv:1603.04856 [hep-th]].
%%CITATION = arXiv:1907.10810%%
}
%\LiTPF
\lref\LiTPF{
  Y.~Z.~Li, Z.~F.~Mai and H.~Lü,
  ``Holographic OPE Coefficients from AdS Black Holes with Matters,''
JHEP {\bf 1909}, 001 (2019).
[arXiv:1905.09302 [hep-th]].
%%CITATION = arXiv:1905.09302%%
}
%\CotlerZFF
\lref\CotlerZFF{
  J.~Cotler and K.~Jensen,
  ``A theory of reparameterizations for AdS$_3$ gravity,''
JHEP {\bf 1902}, 079 (2019).
[arXiv:1808.03263 [hep-th]].
%%CITATION = arXiv:1808.03263%%
}
%\LiZBA
\lref\LiZBA{
  Y.~Z.~Li,
  ``Heavy-light Bootstrap from Lorentzian Inversion Formula,''
[arXiv:1910.06357 [hep-th]].
%%CITATION = arXiv:1910.06357%%
}
%\KologluBCO
\lref\KologluBCO{
  M.~Kologlu, P.~Kravchuk, D.~Simmons-Duffin and A.~Zhiboedov,
  ``Shocks, Superconvergence, and a Stringy Equivalence Principle,''
[arXiv:1904.05905 [hep-th]].
%%CITATION = arXiv:1904.05905%%
}
%\KologluMFZ
\lref\KologluMFZ{
  M.~Kologlu, P.~Kravchuk, D.~Simmons-Duffin and A.~Zhiboedov,
  ``The light-ray OPE and conformal colliders,''
[arXiv:1905.01311 [hep-th]].
%%CITATION = arXiv:1905.01311%%
}
%\DolanDV
\lref\DolanDV{
  F.~A.~Dolan and H.~Osborn,
  ``Conformal Partial Waves: Further Mathematical Results,''
[arXiv:1108.6194 [hep-th]].
%%CITATION = arXiv:1108.6194%%
}
%\KusukiWPA
\lref\KusukiWPA{
  Y.~Kusuki,
  ``Light Cone Bootstrap in General 2D CFTs and Entanglement from Light Cone Singularity,''
JHEP {\bf 1901}, 025 (2019).
[arXiv:1810.01335 [hep-th]].
%%CITATION = arXiv:1810.01335%%
}
%\KusukiGJS
\lref\KusukiGJS{
  Y.~Kusuki and M.~Miyaji,
  ``Entanglement Entropy, OTOC and Bootstrap in 2D CFTs from Regge and Light Cone Limits of Multi-point Conformal Block,''
JHEP {\bf 1908}, 063 (2019).
[arXiv:1905.02191 [hep-th]].
%%CITATION = arXiv:1905.02191%%
}

\Title{
\vbox{\baselineskip8pt
% \hbox{SPIN-07/41} \hbox{
%ITP-UU-07/55}
}}
{\vbox{
\centerline{Multi-stress tensors and next-to-leading}
\vskip.1in
\centerline{singularities in the Regge limit}
}}

\vskip.1in
 \centerline{
Robin Karlsson\footnote{}{karlsson $@$ maths.tcd.ie  }  } \vskip.1in
\centerline{\it 
School of Mathematics, Trinity College Dublin, Dublin 2, Ireland}

\vskip.7in \centerline{\bf Abstract}{
\vskip.2in 
\noindent 
The stress tensor sector of a heavy-heavy-light-light scalar correlator in CFTs with a large central charge and a large gap is defined by the exchange of multi-stress tensor operators. The Regge limit of this correlator is determined by the phase shift of a highly energetic particle propagating in a dual black hole background. Assuming Einstein gravity in the bulk, the phase shift is known perturbatively to all orders in the ratio of the heavy scaling dimension over the central charge. In the CFT, the order counts the number of stress tensors in the multi-stress tensor operator. By Fourier transforming the correlator to position space, the multi-stress tensor contributions to the leading and next-to-leading singularities in the Regge limit are found to all orders in four dimensions. The leading singularity at each order agrees with known results obtained by considering a particle in a dual shockwave background. Moreover, the leading and next-to-leading singularities due to double- and triple-stress tensors with minimal twist are known from lightcone bootstrap and agree with the results derived from the phase shift. 
}
\Date{December 2019}

\listtoc\writetoc
\vskip 1.57in \noindent

%\draftmode

\eject
\newsec{Introduction and summary}
\noindent 
The conformal bootstrap program aims to constrain consistent conformal field theories by studying the implications of conformal symmetry \refs{\FerraraYT,\PolyakovGS}. While originally studied in two dimensions, following developments in \RattazziPE\ this has lead to a better understanding of CFTs in spacetime dimensions $d>2$. In particular, crossing relations between the conformal block expansion of correlation functions in different channels put strong constraints on the CFT data, i.e., the set of scaling dimensions and OPE coefficients. While the crossing relations are in general intractable to solve, imposing extra conditions on the theory such as a large central charge or by studying certain kinematical limits, progress can be made. Simplifications are then typically due to the fact that the expansion in one channel makes certain operators dominant. By crossing symmetry, these has to be reproduced by the expansion in the other channels which typically involve the exchange of infinitely many operators and are, therefore, often complicated. 

One reason for the increased interest in higher-dimensional CFTs is due to its connections to quantum gravity \refs{\MaldacenaRE\WittenQJ-\GubserBC}. Weakly coupled gravitational theories in AdS are related via the AdS/CFT-correspondence to CFTs with large central charge, $C_T$, and large gap, $\Delta_{\rm gap}$, living on the boundary of the spacetime. Restricting to CFTs with this property, much progress has been made in using the bootstrap approach to constrain the CFT data perturbatively in an expansion in $1/C_T$. Especially interesting are gravitational interactions in the bulk; these are related to the exchange of multi-stress tensor operators in the boundary theory, schematically denoted by $[T^k]_{n,l}$. One approach to studying this so-called stress tensor sector in $d>2$ is to consider a heavy-heavy-light-light scalar correlation function, where ``heavy'' and ``light'' refers to operators $\OH$ and $\OL$ with scaling dimension $\DH\sim \OO(C_T)$ and $\DL\sim\OO(1)$ \refs{\KulaxiziDXO\FitzpatrickZQZ\KarlssonQFI\LiTPF\KulaxiziTKD\FitzpatrickEFK-\KarlssonDBD}. Roughly speaking, scaling $\DH$ with the central charge enhances contributions from the stress tensor sector compared to the exchange of generic operators. In $d=2$ the exchange of multi-stress tensors in this correlator are contained in the Virasoro vacuum block \refs{\FitzpatrickVUA\FitzpatrickZHA\HijanoRLA\HijanoQJA\FitzpatrickFOA\CotlerZFF\CollierEXN-\BeskenJYW}, and has been used to study e.g.\ unitary evaporation of black holes in ${\rm AdS}_3$ \refs{\FitzpatrickDLT\FitzpatrickIVE\AnousBHC\FitzpatrickMJQ\ChenNVBIP-\FaulknerHLL} as well as properties of entanglement entropy and quantum chaos \refs{\HartmanMIA\AsplundCOA\CaputaQE\ChenEE\ChenESC\RobertsIFA\KusukiWPA-\KusukiGJS}. Compared to two dimensions where the infinite-dimensional Virasoro algebra puts strong constraints on the stress tensor sector, the higher-dimensional case remains elusive to understand. 

Another approach to making the crossing relations tractable is to study the conformal block expansion in different kinematical limits. Two particularly interesting limits are the lightcone limit and the Regge limit. In the former, the behaviour of the conformal blocks implies that operators with low twist are dominant in the channel $\OL\times\OL\to \OO\to \OH\times\OH$ (direct-channel). This was e.g.\ used in (with both pair of operators being light) \refs{\KomargodskiEK,\FitzpatrickYX} to prove the existence of double-twist operators at large spin. In \refs{\KulaxiziTKD}, the near lightcone heavy-heavy-light-light correlator in large $C_T$ CFTs was studied in $d=4$. There the exchange of minimal-twist ($\tau=4$) double-stress tensors of spin $s=4,6,\ldots,\infty$ was computed. This was extended in \KarlssonDBD\ with a bootstrap approach to determine the exchange of minimal-twist multi-stress tensor operators with arbitrary spin, see also \LiZBA. While the lightcone limit is sensitive to operators with low twist, the Regge limit is dominated by operators with large spin and was originally studied in \refs{\CornalbaXK\CornalbaXM\CornalbaZB\CornalbaAX-\CostaCB}, mainly in the context of AdS/CFT. In particular, the stress tensor sector contains operators of any spin, $[T^k]_{n,l}$ for $k\geq 2$. Studying the Regge limit of this sector is therefore inherently involved. However, in the subsector of minimal-twist multi-stress tensors obtained in \refs{\KulaxiziTKD,\KarlssonDBD} one can approach the Regge limit starting from the explicit resummation of such operators. Utilizing the overlapping regime of validity, the large impact parameter limit of the Regge limit can thus be studied by pure CFT arguments using lightcone bootstrap. 

In this paper, the Regge limit of a scalar heavy-heavy-light-light correlation function in CFTs with large central charge and large gap in $d>2$ (mainly $d=4$) is considered using holography. The Regge limit of a four-point function of pairwise identical light scalar operators is related to the phase shift of $2\to2$ elastic scattering of highly energetic particles at fixed impact parameter in the bulk \refs{\CornalbaXK\CornalbaXM\CornalbaZB\CornalbaAX-\CostaCB}. For further discussion about the Regge limit and the phase shift in holographic CFTs see also \refs{\CostaTWZ\LiLMH\KulaxiziIXA-\CamanhoAPA}. In the heavy-heavy-light-light case, the phase shift \KulaxiziDXO\ was defined in the bulk in terms of the Shapiro time delay and the angle deflection of a highly energetic particle propagating in an AdS-Schwarzschild background; it was then calculated to all orders in a perturbative expansion in the Schwarzschild radius in units of the AdS radius. In the CFT the phase shift is related to a Fourier transform of the correlator and the expansion parameter is given by $\mu\sim {\DH\over C_T}$. At $k$-th order, the phase shift is given by a massive scalar propagator in $(k(d-1)-(k-1))$-dimensional hyperbolic space. On the other hand, the leading Regge behaviour of a conformal block in $d$ dimensions takes the form of a scalar propagator in $(d-1)$-dimensional hyperbolic space. The higher-dimensional propagators appearing in the phase shift can, however, be decomposed into infinite sums of propagators with increasing scaling dimensions in $H^{d-1}$ \KarlssonQFI. This appears to be a more natural representation of the phase shift from the boundary point of view.

In particular, we will study the leading and next-to-leading singularities of the stress tensor sector of the heavy-heavy-light-light correlator in the Regge limit. This is done perturbatively in $\mu$ and the stress tensor sector of the correlator $\GG(z,\zbar)$ (to be defined below) is given by
\eqn\stressSectInt{\eqalign{
  \GG(\sigma,\rho) = \sum_{k=0}^\infty \mu^k\GG^{(k)}(\sigma,\rho),\qquad &\sigma e^{\rho}=1-z\cr
  &\sigma e^{-\rho}=1-\zbar.
}}
The stress tensor sector $\GG$ of the correlator contains the contribution of multi-stress tensor operators in the direct-channel expansion of the correlator $\OL\times\OL\to\mu^k[T^k]_{n,l}$. Here it is seen that the contribution at $k$-th order is due to multi-stress tensors made out of $k$ stress tensors. 

At $k$-th order, the stress tensor sector $\GG^{(k)}$ behaves as follows in the Regge limit:
\eqn\RegLim{
  \GG^{(k)}(\sigma,\rho) = {F_{k,{\rm L}}(\rho)\over \sigma^k}+{F_{k,{\rm NL}}(\rho)\over\sigma^{k-1}}+\OO(\sigma^{-k+2})\qquad \quad \sigma\to 0,\quad \rho{\rm-fixed},
}
for some functions $F_{k,{\rm L}}(\rho)$ and $F_{k,{\rm NL}}(\rho)$. We define the leading and next-to-leading Regge singularity of the stress tensor sector of the correlator $\GG^{(k)}$ at $\OO(\mu^k)$ by 
\eqn\defReggeSing{\eqalign{
  &{\rm Leading\,\,Regge\,\,singularity:} \qquad {F_{k,{\rm L}}(\rho)\over \sigma^k},\cr
  &{\rm Next{\rm-}to{\rm-}leading\,\,Regge\,\,singularity:} \qquad {F_{k,{\rm NL}}(\rho)\over \sigma^{k-1}}.
}}
The aim of this paper is to calculate $F_{k,{\rm L}}$ and $F_{k,{\rm NL}}$ for any value of $k$ and fixed $\rho$. This is done by Fourier transforming the momentum space correlator given in terms of the bulk phase shift.

\subsec{Summary of results}
We consider a heavy-heavy-light-light correlator in a four-dimensional CFT with large central charge and large gap. The stress tensor sector $\GG(z,\zbar)$ of the heavy-heavy-light-light correlator is defined by the exchange of multi-stress tensors:
\eqn\DefStressTensor{
  \GG(z,\zbar) = {1\over \GG_{0}(z,\zbar)}\lim_{x_4\to\infty} x_4^{2\DH}\langle \OH(x_4)\OL(1)\OL(z,\zbar)\OH(0)\rangle \Big|_{\rm multi-stress\; tensors},
}
with $\GG_0(z,\zbar)$ the disconnected correlator. The restriction to multi-stress tensor exchanges refers to considering exchanges in the direct-channel of the following form:
\eqn\gStressMultDef{
  \OL\times\OL\to\mu^k [T^k]_{n,l}\to \OH\times\OH,
} 
where $\mu\sim {\DH\over C_T}$ is kept fixed and its power counts the number of stress tensors in the multi-stress tensor operator $[T^k]_{n,l}$, with $n=0,1,\ldots$ and $l=0,2,4,\ldots$.

We consider the Regge limit defined by 
\eqn\ReggelimIntro{\eqalign{
  {\rm Regge\,\, limit:}\qquad &z\to e^{-2\pi i}z\qquad {\rm with}\quad \sigma\to 0,\quad \rho{\rm-fixed},\cr
    &\sigma e^\rho = 1-z\qquad \sigma e^{-\rho}=1-\zbar.
}}  
In this limit, the momentum space correlator $\BB(S,L)$ is assumed to be given by the exponentiation of the bulk phase shift $\delta(S,L;\mu)$, where $S$ is the energy and $L$ the impact parameter:
\eqn\momSpace{
  \BB(S,L)=\BB_0(S)e^{i\delta(S,L;\mu)}. 
}

The phase shift $\delta(S,L;\mu)$ was calculated in Einstein gravity in \KulaxiziDXO\ to all orders in $\mu$ and we denote the $k$-th term in that expansion $\delta^{(k)}$. In the Regge limit $S\gg 1$, the phase shift is linear in $S$ and the leading ($\sim\sigma^{-k}$) and next-to-leading Regge singularities ($\sim\sigma^{-k+1}$) of $\GG^{(k)}(\sigma,\rho)$ are due to terms in \momSpace\ of the form:
\eqn\import{
  \BB(S,L)\Big|_{\mu^k} = \BB_0(S)\Big[{(i\delta^{(1)})^k\over k!}+i\delta^{(2)}{(i\delta^{(1)})^{k-2}\over (k-2)!}+\ldots\Big],
} 
where the ellipses denote terms that contribute at subleading order in $\sigma\to0$. 

By Fourier transforming the first term in the brackets in \import, it is found that the leading Regge singularities of $\GG$ are given by 
\eqn\LeadingIntro{\eqalign{
  \GG^{(k)}(\sigma,\rho)\Big|_{\sigma^{-k}}= \sum_{n=0}^{\infty}&{(3i\pi)^k\over k!}{n+k-2\choose n}(\DL)_{2k+n-1}(\DL-1)_{1-k-n}\cr
  &\times{e^{-(3k+2n)\rho}\over \sigma^{k}(1-e^{-2\rho})},
}}
for $k=1,2,\ldots.$ This agrees with the result in \FitzpatrickEFK\ obtained by considering a light particle propagating in a shockwave background. In this case, we see that the leading Regge singularities are fully determined by the phase shift at first order in $\mu$. The first-order phase shift is in turn fixed by the exchange of the stress tensor in the CFT and is therefore universal. 

The next-to-leading Regge singularity $\sim \sigma^{-k+1}$ at $\OO(\mu^k)$ gets two contributions, there is a subleading correction in $\sigma$ coming from $(\delta^{(1)})^k$ in \import\ as well as a contribution from $\delta^{(2)}(\delta^{(1)})^{k-2}$ in \import. The former gives the following contribution:
\eqn\ggas{\eqalign{
  \GG^{(k)}(\sigma,\rho)\Big|_{(\delta^{(1)})^k,\sigma^{-k+1}} =&{1\over2}\sum_{n=0}^{\infty}{(3i\pi)^k\over k!}{n+k-2\choose n}(\DL)_{2k+n-1}(\DL-1)_{1-k-n}\cr
&\times\Big[(k+n-1)e^{-\rho}-(2k+n)e^\rho\Big]{e^{-(3k+2n)\rho}\over \sigma^{k-1}(1-e^{-2\rho})},
}}
while the latter gives:
\eqn\nextToLead{\eqalign{
  \GG^{(k)}(\sigma,\rho)\Big|&_{\delta^{(2)}(\delta^{(1)})^{k-2},\sigma^{-k+1}}= {1\over 320}\sum_{p=0}^\infty\sum_{n=0}^p {i^{k-1}(3\pi)^{k-2}\over (k-2)!}{k+p-n-3\choose p-n}\cr
  &\times (\DL)_{2k+p-2}(\DL-1)_{1-k-p}\lambda_2(n)\bar{c}_{6+2n,2}{e^{-(3k+2p-1)\rho}\over \sigma^{k-1}(1-e^{-2\rho})},
}}
where $\bar{c}_{6+2n,2}$ are constants given in (2.17). The coefficients $\lambda_2(n)$ are related to the decomposition of the second-order phase shift into Regge conformal blocks in (4.3) and are valid assuming there are no higher-derivative corrections in the bulk gravitational action. Adding \ggas\ and \nextToLead\ gives the full expression for the next-to-leading singularities $\sim\sigma^{-k+1}$ at $\OO(\mu^k)$ due to multi-stress tensor of the schematic form $[T^k]_{n,l}$.

In particular, in the large impact parameter limit $\rho\to\infty$, only the $p=0$ term in \nextToLead\ contributes
\eqn\nextToLeadLargeRho{\eqalign{
  \GG^{(k)}(\sigma,\rho)\Big|_{\delta^{(2)}(\delta^{(1)})^{k-2},\sigma^{-k+1}}\rhoequal {35\over 6}{(3i\pi)^{k-1}\over (k-2)!}(\DL)_{2k-2}(\DL-1)_{1-k}{e^{-(3k-1)\rho}\over \sigma^{k-1}}.
}}
The result at finite impact parameter in \nextToLead\ is valid assuming Einstein gravity in the bulk. In the large impact parameter regime given in \nextToLeadLargeRho, the result is expected to be universal since there is by now much evidence of universality in the minimal-twist subsector of multi-stress tensors \refs{\FitzpatrickZQZ,\FitzpatrickEFK,\HuangFOG,\KarlssonDBD}. We find perfect agreement between \LeadingIntro-\nextToLead\ and known results for minimal-twist double-stress and triple-stress tensors obtained using lightcone bootstrap \refs{\KulaxiziTKD,\KarlssonDBD}.

In Section 2, general properties of the heavy-heavy-light-light correlator in CFTs with large central charge is considered as well as its connection to the bulk phase shift. In Section 3, the procedure for decomposing products of Regge conformal blocks in $d=2,4$ is described. In Section 4, the leading and next-to-leading Regge singularities in four dimensions are found from the exponentiation of the phase shift. Section 5 is devoted to discussion and the appendices contain some technical details and further matching with results obtained from lightcone bootstrap. 

\newsec{Heavy-heavy-light-light correlator in holographic CFTs}
This section contains a review of the stress tensor sector of a heavy-heavy-light-light scalar four-point function in CFTs with a large central charge, $C_T$, and a large gap, $\Delta_{\rm gap}$. This sector consists of multi-stress tensors labeled by their twist and spin. In the lightcone limit, multi-stress tensors with minimal twist and arbitrary spin dominate. Since the twist is bounded from below, one can study the correlator perturbatively in a kinematical expansion close to the lightcone. On the other hand, in the Regge limit multi-stress tensors of highest spin dominate due to the behaviour $\sigma^{1-J}$ of the blocks in the Regge limit $\sigma\to0$, with $J$ being the spin. This limit is therefore difficult to study a priori in CFTs. Instead, we use the bulk phase shift calculated in the dual gravitational theory to extract contributions to the stress tensor sector of the correlator in the Regge limit. Approaching the large impact parameter limit of the Regge limit, we can make contact with results obtained using lightcone bootstrap.

In Section 2.1, the heavy-heavy-light-light correlator in CFTs is reviewed with emphasis on its behaviour in the lightcone- and the Regge limit. In Section 2.2, known results for the subsector of minimal-twist double- and triple-stress tensors are studied in the large impact parameter regime of the Regge limit. In Section 2.3, the connection between the bulk phase shift and the heavy-heavy-light-light correlator is explained following \KulaxiziDXO.  

\subsec{Review of the heavy-heavy-light-light correlator in CFTs with large central charge}
The stress-tensor sector of a heavy-heavy-light-light four-point function of pairwise identical scalars $\langle \OH(x_4)\OL(x_3)\OL(x_2)\OH(x_1)\rangle$ is defined by
\eqn\defStressSector{
  \GG(z,\zbar) = {1\over\GG_{0}(z,\zbar)}\lim_{x_4\to\infty} x_4^{2\DH}\langle \OH(x_4)\OL(1)\OL(z,\zbar)\OH(0)\rangle \Big|_{\rm multi-stress\; tensors},
} 
with $z,\zbar$ the cross-ratios 
\eqn\cross{\eqalign{
  u &= (1-z)(1-\zbar) = {x_{14}^2 x_{23}^2\over x_{13}^2x_{24}^2},\cr
  v &= z\zbar = {x_{12}^2 x_{34}^2 \over x_{13}^2 x_{24}^2},
}}
and $\GG_0(z,\zbar)=[(1-z)(1-\zbar)]^{-\DL}$ is the disconnected correlator. The precise meaning of the stress tensor sector is described below.
``Heavy'' in this case refers to taking the scaling dimension $\DH$ of $\OH$ to be $\OO(C_T)$, with $C_T\to\infty$ being the central charge. The scaling dimension $\DL$ of the ``light'' operator $\OL$ remains $\OO(1)$ in this limit. The multi-stress tensor sector is then studied perturbatively in $\mu$
\eqn\pertG{
  \GG(z,\zbar) = \sum_{k=0}^\infty \mu^k\GG^{(k)}(z,\zbar),
}
with
\eqn\defMu{
  \mu := {4\Gamma(d+2)\over(d-1)^2 \Gamma({d\over 2})^2}{\DH\over C_T}.
} 

The notion of the stress tensor sector comes from the conformal block expansion in the ``T-channel'' (direct-channel) $\OL(1)\times\OL(z,\zbar)\to\lambda_{\OL\OL\OO_{\Delta,s}}\OO_{\Delta,s}$
\eqn\tCh{
  \GG(z,\zbar) = \sum_{\OO_{\Delta,s}}P^{(HH,LL)}_{\OO_{\Delta,s}} g_{\Delta,s}^{}(1-z,1-\zbar),
}
where $g_{\Delta,s}(1-z,1-\zbar)$ is the conformal block for a primary operator with dimension $\Delta$ and spin $s$, and we have defined the product of OPE coefficients by 
\eqn\prodOPE{
  P^{(HH,LL)}_{\OO_{\Delta,s}} := \left(-{1\over 2}\right)^s\lambda_{\OH\OH\OO_{\Delta,s}}\lambda_{\OL\OL\OO_{\Delta,s}}.
}

At $\OO(\mu^k)$ in the T-channel expansion \tCh, ``multi-stress tensor'' operators which we denote by $[T^k]^{(i)}_{n,l}$ contribute. For $k=1$ this is the exchange of the stress tensor $T_{\mu\nu}$ with $P^{(HH,LL)}_{T_{\mu\nu}}$ determined by Ward identities
\eqn\opeStress{
  P^{(HH,LL)}_{T_{\mu\nu}} = \mu{\DL\over 4}{\Gamma({d\over 2}+1)^2\over \Gamma(d+2)}.
}

Naively, at $\OO(\mu^2)$ there are three infinite families of double-stress tensors of the following schematic form:
\eqn\doubleStress{\eqalign{
  [T^2]_{n,l}^{(0)}=&:T_{\mu\nu}(\pa^2)^n\pd_{\mu_1}\ldots \pd_{\mu_l}T_{\rho\kappa}:,\cr
  [T^2]_{n,l}^{(1)}=&:T_{\mu\rho}(\pa^2)^n\pd_{\mu_1}\ldots \pd_{\mu_l}{T^\rho}_{\nu}:,\cr
  [T^2]_{n,l}^{(2)}=&:T_{\rho\kappa}(\pa^2)^n\pd_{\mu_1}\ldots \pd_{\mu_l}T^{\rho\kappa}:,
}}
where the superscript denotes the number of contracted pair of indices between stress tensors and $n=0,1,2,\ldots$ and $l=0,2,4,\ldots$. The double-stress tensors in \doubleStress\ have the following twist ($\tau=\Delta-s$) $\tau^{(2,i)}_{n,l}$ and spin $s^{(2,i)}_l$:
\eqn\dataDoubleStress{\eqalign{
  \tau^{(2,0)}_{n,l} &= 4+2n \qquad s^{(2,0)}_l=4+l,\cr
  \tau^{(2,1)}_{n,l} &= 6+2n \qquad s^{(2,1)}_l=2+l,\cr
  \tau^{(2,2)}_{n,l} &= 8+2n \qquad s^{(2,2)}_l=l.\cr
}}
From \dataDoubleStress\ it is seen that the operators in \doubleStress\ have the same quantum numbers for suitable values of $n,l$ and therefore possibly mix among each other. At $\OO(\mu^k)$, for $k>1$, there again exist different multi-stress tensor operators with overlapping quantum numbers, similar to the double-stress tensor case in \doubleStress. 

The T-channel conformal blocks in $d=2,4$ are given by \refs{\DolanHV,\DolanUT}:
\eqn\TChannelBlock{\eqalign{
  g^{d=2}_{\Delta,J}(1-z,1-\zbar) &= f_{\Delta+J\over 2}(z)f_{\Delta-J\over 2}(\zbar)+(z\lra\zbar),\cr
  g^{d=4}_{\Delta,J}(1-z,1-\zbar) &= {(1-z)(1-\zbar)\over \zbar-z}\Big[f_{\Delta+J\over 2}(z)f_{\Delta-J-2\over 2}(\zbar)-(z\lra\zbar)\Big],
}}
with $f_a(z)$ a $SL(2,R)$ conformal block
\eqn\fFunc{
  f_a(z) = (1-z)^a{}_2F_1(a,a;2a;1-z).
}

We are interested in the lightcone- and the Regge limit defined respectively by:
\eqn\kinematics{\eqalign{
  &{\rm Lightcone\,\, limit:}\qquad \zbar\to 1,\, z\,\,{\rm fixed},\cr
  &{\rm Regge\,\, limit:}\qquad ze^{-2\pi i}\to 1,\, {1-z\over 1-\zbar}\,\,{\rm fixed},
}}
wherein the Regge limit $ze^{-2\pi i}$ denotes analytic continuation of $z$ clockwise around $z=0$. It is further convenient to introduce coordinates $(\sigma,\rho)$ in the Regge limit (after analytic continuation):
\eqn\sigmaAndRho{\eqalign{
  \sigma e^{\rho} = 1-z\cr
  \sigma e^{-\rho} = 1-\zbar, 
}}
in which the Regge limit corresponds to $\sigma\to 0$ with $\rho$ fixed. We further define the subsequent limit $\rho\to\infty$ as the large impact parameter limit. The latter regime can be reached from the lightcone limit by analytically continuing $z\to e^{-2\pi i}z$ and taking $z\to 1$. 

In the lightcone limit, the blocks $g_{\Delta,J}(1-z,1-\zbar)$ are given to leading order in $1-\zbar$ in any dimension by:
\eqn\twistExp{
  g_{\Delta,J}(1-z,1-\zbar) \lzbarone (1-\zbar)^{\tau\over 2}f_{\Delta+J\over 2}(z), 
}
where $\tau=\Delta-J$ is the twist and by $\lzbarone$ we keep only the leading term in the expansion. Thus it is seen that in the lightcone limit operators with low twist will dominate. From \dataDoubleStress, it is clear that the multi-stress tensor operators with the lowest twist are those with $n=0$ and no contractions.

It was argued in \KarlssonDBD\ that the subsector of minimal-twist multi-stress tensor operators is universally fixed by crossing symmetry in terms of the exchange of the stress tensor. The contribution of minimal-twist multi-stress tensors $[T^k]_{0,l}^{(0)}$ to the heavy-heavy-light-light correlator takes the following particular form:  
\eqn\gkhd{
  \GG_{\rm LC}^{(k)}(z,\zbar) = (1-\zbar)^{k({{d-2}\over 2})}\sum_{\{i_p\}}a_{i_1\ldots i_k} f_{i_1}(z)\ldots f_{i_k}(z), \qquad \sum_{p=1}^k i_p = k\left({{d+2}\over 2}\right).
}
Here the subscript LC denotes that this is the leading contribution of the stress tensor sector in the lightcone limit. The coefficients $a^{(d=4)}_{i_1i_2}$ were first computed in \KulaxiziTKD\ due to the exchange of minimal-twist double-stress tensors in four dimensions. In \KarlssonDBD, a bootstrap procedure to calculate $a_{i_1\ldots i_k}$ for any $k$ in even dimensions was given. Explicitly, the contribution from minimal-twist double-stress tensors in $d=4,6$ were found as well as that from minimal-twist triple-stress tensors in $d=4$. Note that each term $\GG^{(k)}_{\rm LC}$ sums an infinite number of multi-stress tensor operators $[T^k]_{0,l}$ with twist $k(d-2)$ and spin $2k+l$, for $l=0,2,\ldots$. For more details on the minimal-twist multi-stress tensors, see \KarlssonDBD.

The leading behaviour of the block $g^{\circlearrowright}_{\Delta,J}(1-z,1-\zbar)$ in the Regge limit, $\sigma\to 0$, is given by 
\eqn\reggeBlock{\eqalign{
  g^{\circlearrowright,d=2}_{\Delta,J}(\sigma,\rho) &= i\bar{c}_{\Delta,J}{e^{-(\Delta-1)\rho}\over \sigma^{J-1}}+\ldots,\cr
  g^{\circlearrowright,d=4}_{\Delta,J}(\sigma,\rho) &= i\bar{c}_{\Delta,J}{e^{-(\Delta-1)\rho}\over (1-e^{-2\rho})\sigma^{J-1}}\cr
  &\times\Big[1-{\sigma\over 4}\Big((\Delta+J-2)e^{\rho}+(2+J-\Delta)e^{-\rho}\Big)+\OO(\sigma^2)\Big]
}}
with
\eqn\cBarDef{
  \bar{c}_{\Delta,J} = {4^{\Delta+J-1} \Gamma\left({\Delta+J-1\over 2}\right)\Gamma\left({\Delta+J+1\over 2}\right)\over \Gamma({\Delta+J\over 2})^2}.
}
Here we have included the first subleading correction in $\sigma\to0$ in four dimensions since this will be needed later on.
More generally, the leading behaviour in the Regge limit in any dimension is given by, see e.g.\ \refs{\DolanDV,\KulaxiziIXA},
\eqn\BlockGenerald{
  g^{\circlearrowright}_{\Delta,J}(\sigma,\rho) = ic_{\Delta,J}\sigma^{1-J}\Pi_{\Delta-1,d-1}(\rho)+\ldots,
}
where $\Pi_{\Delta-1,d-1}(\rho)$ is $(d-1)$-dimensional hyperbolic space propagator of a particle with mass-squared $m^2=(\Delta-1)^2$
\eqn\HyperbolicPropagator{
  \Pi_{\Delta-1,d-1}(\rho) = {\pi^{1-{d\over 2}}\Gamma(\Delta-1)\over 2\Gamma(\Delta-{d-2\over 2})}e^{-(\Delta-1)\rho}{}_2F_1\Big({d-2\over 2},\Delta-1;\Delta-{d-2\over 2};e^{-2\rho}\Big)
}
and
\eqn\Coefficient{
c_{\Delta, J}={4^{\Delta+J-1} \Gamma\left({\Delta+J-1\over 2}\right)\Gamma\left({\Delta+J+1\over 2}\right)\over \Gamma({\Delta+J\over 2})^2} \, {2\Gamma\left(\Delta-{d\over 2}+1\right)\over \pi^{1-{d\over 2}} \Gamma\left({\Delta-1}\right)}.
}

\subsec{The Regge limit of minimal-twist double- and triple-stress tensors}
The contribution of minimal-twist operators in \gkhd\ was obtained by solving crossing relations in a large $C_T$ CFT; a holographic dual was not assumed. In this paper we assume a large $C_T$ CFT with large $\Delta_{\rm gap}$. Explicitly, we use the bulk phase shift calculated in the gravitational dual to study the CFT correlator in the Regge limit. To make contact between the Regge limit and the lightcone limit, we analytically continue $\GG_{\rm LC}^{(k)}(z,\zbar)$ according to $z\to e^{-2\pi i}z$:
\eqn\defLCAnalytCont{
  \GG_{\rm LC}^{(k),\circlearrowright}(z,\zbar):= \GG_{\rm LC}^{(k)}(ze^{-2\pi i},\zbar).
} 
Sending also $z\to1$, this is the large impact parameter limit of the Regge limit. 

Using the explicit results for $\GG^{(2)}_{\rm LC}$ in $d=4$ \KulaxiziTKD, we find the following leading and next-to-leading singularities in the large impact parameter limit due to minimal-twist double-stress tensors
\eqn\ratioCorrSecond{\eqalign{
   \GG^{(2),\circlearrowright}_{\rm LC}(\sigma,\rho) &= -{9\pi^2\DL(\DL+1)(\DL+2)\over 2(\DL-2)}{e^{-6\rho}\over \sigma^2}+\cr
   &+\Big[{35i\pi\DL(\DL+1)\over 2(\DL-2)}+{18\pi^2\DL(\DL+1)(\DL+2)\over 2(\DL-2)}\Big]{e^{-5\rho}\over \sigma}+\ldots,
}} 
where the ellipses denote non-singular terms as $\sigma\to 0$. Likewise, in the large impact parameter limit of $\GG^{(3),\circlearrowright}_{\rm LC}$ due to the exchange of minimal-twist triple-stress tensors \KarlssonDBD, one finds the following leading and next-to-leading singularities in the Regge limit:
\eqn\ratioCorrThird{\eqalign{
  \GG^{(3),\circlearrowright}_{\rm LC}(\sigma,\rho)\Big|_{\sigma^{-3}} &= -{9i\pi^3\DL(\DL+1)(\DL+2)(\DL+3)(\DL+4)\over 2(\DL-2)(\DL-3)}{e^{-9\rho}\over \sigma^3},\cr
  \GG^{(3),\circlearrowright}_{\rm LC}(\sigma,\rho)\Big|_{\sigma^{-2}} &= \Big[-{105\pi^2\DL(\DL+1)(\DL+2)(\DL+3)\over 2(\DL-2)(\DL-3)}\cr
  &+{27i\pi^3\DL(\DL+1)(\DL+2)(\DL+3)(\DL+4)\over 2(\DL-2)(\DL-3)}\Big]{e^{-8\rho}\over\sigma^2}.
}}
The results \ratioCorrSecond-\ratioCorrThird\ from lightcone bootstrap will be compared to the results obtained using the bulk phase shift to study the Regge limit. While we are mainly interested in terms that behave as $\sigma^{-k}$ and $\sigma^{-k+1}$ at $\OO(\mu^k)$ in the Regge limit, the term proportional to $\sigma^{-1}$ at $\OO(\mu^3)$ is further given by:
\eqn\ratioCorrThirdsigmaOne{\eqalign{
  \GG^{(3),\circlearrowright}_{\rm LC}(\sigma,\rho)\Big|_{\sigma^{-1}} &= \Big[i\Big({1155\pi\DL(\DL+1)(\DL+2)\over 8(\DL-2)(\DL-3)}\cr 
  &-{9\pi^3\DL(\DL+1)(\DL+2)(\DL+4)(19+7\DL)\over 4(\DL-2)(\DL-3)}\Big)\cr
  &+{525\pi^2\DL(\DL+1)(\DL+2)(\DL+3)\over 4(\DL-2)(\DL-3)}\Big]{e^{-7\rho}\over \sigma}.
}}

\subsec{Bulk phase shift of a light particle in an AdS black hole background}
The relationship between the bulk phase shift and the heavy-heavy-light-light correlator was described in \KulaxiziDXO\ which we briefly review here for completeness. We consider a four-point function defined on the cylinder parameterized by time $\tau$ and a unit vector $\hat{n}$ on $S^{d-1}$:
\eqn\fourptcylinder{
  G(x)\equiv \langle \OO_H^{\rm cyl}(\tau_4,\hat{n}_4)\OO_L^{\rm cyl}(\tau_3,\hat{n}_3)\OO_L^{\rm cyl}(\tau_2,\hat{n}_2)\OO_H^{\rm cyl}(\tau_1,\hat{n}_1)\rangle. 
}
Inserting the heavy operators at $\tau_{4,1}=\pm\infty$ and going to the plane using the transformation $r=e^\tau$, we have
\eqn\fourptPlaneDef{
  G(x) = (r_2r_3)^\DL {\hat{\GG}(z,\zbar)\over x_{32}^{2\DL}},
}
where the cross-ratios are given by 
\eqn\crossRatiosGravity{\eqalign{
  z\zbar &= {x_2^2\over x_3^2} = e^{2(\tau_2-\tau_3)}\cr
  (1-z)(1-\zbar) &= {x_{23}^2\over x_3^2} = 1+e^{2(\tau_2-\tau_3)}-2e^{\tau_2-\tau_3}\hat{n}_2\cdot\hat{n}_3.     
}}
The function $\hat{\GG}(z,\zbar)$ can be expanded in conformal blocks on the plane. Especially, we will be interested in the stress tensor sector $\GG$ of $\hat{\GG}$:
\eqn\gStressSector{
  \GG(z,\zbar)\equiv \hat{\GG}(z,\zbar)\Big|_{\rm multi-stress\; tensors}
}
As we will see, when Fourier transforming the phase shift, there are contributions to $\hat{\GG}(z,\zbar)$ coming from double-trace operators that are of the schematic form $[\OL\OL]_{n,l}$. By  definition, these do not contribute to $\GG(z,\zbar)$

We introduce two points $P_2$ and $P_3$ on the cylinder which differ by Lorentzian time $\pi$ and are diametrically opposite on the sphere, i.e.\ $\hat{n}(P_3)=-\hat{n}(P_2)$. By translational and rotational invariance, the operator $\OO_L(x_2)$ is inserted at $P_2$ and $\OO_L(x_3)$ is inserted close to $P_3$ with $\hat{n}_3\cdot\hat{n}(P_3)=\cos\varphi$. Starting from Euclidean kinematics, we Wick-rotate by $\tau_i\to it_i$ and set $t_3-t_2=\pi+x_0-i{\epsilon\over2}$, where $x^0\geq 0$ parameterizes the time delay. Using \crossRatiosGravity\ one can solve for $z,\zbar$ in terms of $x^\pm = x^0\pm\varphi$: 
\eqn\crossGravity{\eqalign{
  z &= e^{-ix^+}\cr
  \zbar &= e^{-ix^-}.
}}
Note that a highly energetic light particle in pure AdS starting at $P_2$ will propagate to the point $P_3$; the $(x^0,\varphi)$-coordinates measure the position of $\OO_L(x_3)$ relative to the point $P_3$. These kinematics are obtained starting with the operators close to $P_2$, corresponding to $x^+\approx -2\pi$, and then $\OL(\tau_3,\hat{n}_3)$ is moved close to $P_3$ by taking $x^+\to x^++2\pi$. In terms of the cross-ratios, this corresponds to taking $z\to e^{-2\pi i}z$. With these kinematics, the correlator $G(x)$ in \fourptPlaneDef\ is given in the Regge limit $x^{\pm}\to 0$, with their ratio kept fixed, by
\eqn\gCylinder{
  G(x)= {\hat{\GG}(z,\zbar)\over(-x^2-i\epsilon x^0)^\DL}\left[1+\OO((x^+)^2)\right],
}
with $-x^2=(x^0)^2-\varphi^2$.

The phase shift is defined by the following Fourier transform:
\eqn\DefPhaseShift{
  \BB(p)\equiv \BB_0(p)e^{i\delta}=\int d^d x G(x)e^{-ipx},
}
where $\BB_0(p)$ denotes the Fourier transform of the disconnected correlator and $e^{i\delta}$ contain the (non-trivial) dynamics of the correlator. In the language of the dual CFT, the phase shift encodes the Fourier transform of the conformal block expansion and the corresponding OPE coefficients of non-trivial operators in the direct-channel. Explicitly, the Fourier transform of the disconnected correlator is given by
\eqn\TransfDisc{
  \BB_0(p)= \int d^d x {e^{-ipx}\over (-x^2-i\epsilon x^0)^\DL} = \theta(p^0)\theta(-p^2)e^{i\pi\DL}C(\DL)(-p^2)^{\DL-{d\over 2}}
}
where 
\eqn\cConst{
  C(\DL) = {2^{d+1-2\DL}\pi^{1+{d\over 2}}\over \Gamma(\DL)\Gamma(\DL-{d-2\over 2})}. 
}
Here the combination $\theta(p^0)\theta(-p^2)$ ensures that $p$ lies in the upper Milne wedge $M^+$.

We further introduce the parametrization $p^\mu=\sqrt{-p^2}\omega^\mu$ in terms of two vectors $\omega$ and $\bar{e}$, such that $\omega^2=\bar{e}^2=-1$ and $\bar{e}^0=1$ with all other components set to $0$. Then 
\eqn\SandRho{\eqalign{
  S &= \sqrt{-p^2},\cr
  \cosh L &= -\ebar\cdot \omega = {p^++p^-\over 2\sqrt{-p^2}}.
}}
Likewise, we define $x^\mu=\sqrt{-x^2}e^\mu$ with $e^2=-1$ such that
\eqn\coord{\eqalign{
  \sqrt{-x^2}&=\sqrt{-\log z\log\zbar},\cr
  -e\cdot\bar{e} &=  {i\log z\zbar\over 2\sqrt{-\log z\log \zbar}}.
}}
Expanding \coord\ in the Regge limit $\sigma\to0$, one finds  
\eqn\coordCorr{\eqalign{
  \sqrt{-x^2} &= -i\sigma\Big(1+{\sigma\over2}\cosh\rho+\ldots\Big),\cr
  -e\cdot\bar{e} &= \cosh\rho+{e^{-2\rho}(1-e^{\rho})^2(1+e^{\rho})^2\over 8}\sigma+\ldots,
}} 
where the ellipses denote subleading corrections in $\sigma$.

\newsec{Fourier transforming products of Regge conformal blocks}
Following \KulaxiziDXO, we review how a single Regge conformal block in momentum space can be transformed into position space in any dimension. The leading result in $\sigma\to0$ can then be identified with the leading Regge behaviour of a conformal block due to an operator exchange in the direct-channel. In the case when the operator appears in the spectrum, its coefficient is related to the product of OPE coefficients\foot{The term ``effective operator'' is used below when the Fourier transform of a Regge conformal block can be identified with the leading Regge behaviour of a conformal block even though such an operator does not appear in the expansion \tCh.}. 

In Section 3.1, we show that a product of Regge conformal blocks in two dimensions is again a Regge conformal block. In Section 3.2, the four-dimensional case is considered where, on the other hand, it is shown that products of Regge conformal blocks can be decomposed into an infinite sum of Regge conformal blocks of different twist $\Delta-J$. Using this decomposition, it is straightforward to do the Fourier transform and read off the contribution to the position space correlator. In particular, in the large impact parameter limit $\rho\gg 1$, only the term with minimal twist in the decomposition is important. In this limit one can, therefore, approximate products of Regge conformal blocks in $d=4$ with a single Regge conformal block. This is reminiscent of what happens in $d=2$.  

A Regge conformal block was defined in \KarlssonQFI\ by
\eqn\ReggeBlock{
  g_{\Delta,J}^{\rm R}(S,L) = ic_{\Delta,J}S^{J-1}\Pi_{\Delta-1,d-1}(L),
} 
with $\Pi_{\Delta-1,d-1}(L)$ a $(d-1)$-dimensional hyperbolic space propagator of a particle with mass-squared $m^2=(\Delta-1)^2$, defined in \HyperbolicPropagator, and $c_{\Delta,J}$ given by \Coefficient. Note that the Regge conformal blocks in \ReggeBlock\ is identical to the leading Regge behaviour of the analytically continued blocks in \BlockGenerald\ with the following replacement $S\to\sigma^{-1}$ and $L\to\rho$.

The hyperbolic space propagator in \HyperbolicPropagator\ can be written in terms of functions $\Omega_{i\nu}={i\nu\over 2\pi}(\Pi_{i\nu+{d\over 2}-1}-\Pi_{-i\nu+{d\over 2}-1})$ as\foot{For brevity, we denote $\Pi_{\Delta-1}\equiv \Pi_{\Delta-1,d-1}$ and likewise for $\Omega_{i\nu}$.}
\eqn\identityOmegaPi{
  \Pi_{\Delta-1}(L) = \int^{\infty}_{-\infty} d\nu {\Omega_{i\nu}(L)\over \nu^2+(\Delta-{d\over 2})^2},
}
which can be shown using \HyperbolicPropagator\ and deforming the integration contour to compute the integral. The functions $\Omega_{i\nu}$ constitute a basis of regular eigenfunctions of the Laplacian operator on $H_{d-1}$, for more details, see e.g.\ \PenedonesNS. 

Consider the contribution to the correlator due to a single Regge conformal block of dimension $\Delta$ and spin $J$:
\eqn\corrMom{
	\BB(S,L)\Big|_{\Delta,J}=\BB_0(S)\lambda g^{\rm R}_{\Delta,J}(S,L),
}
where $\lambda$ is a numerical coefficient and $\BB_0(S)$ is the disconnected correlator given in \TransfDisc. The position space result from \corrMom\ is given by the Fourier transform
\eqn\dfs{
	G(x)\Big|_{\Delta,J}= \lambda\int_{M^+} {d^dp\over (2\pi)^d} e^{ipx}\BB_0(S)g^{\rm R}_{\Delta,J}(S,L),
}
which by inserting \ReggeBlock\ and using \identityOmegaPi\ can be written as 
\eqn\afd{
	G(x)\Big|_{\Delta,J}= ic_{\Delta, J}\lambda\int_{M^+} {d^dp\over (2\pi)^d} e^{ipx}\BB_0(S)S^{J-1}\int^{\infty}_{-\infty} d\nu {\Omega_{i\nu}(\omega\cdot\bar{e})\over \nu^2+(\Delta-{d\over 2})^2}.
}
We then need the following identity derived in \KulaxiziDXO: 
\eqn\FourierIdentity{
  {2^{1-a}e^{i\pi a\over 2}\over \pi^{d-2\over 2}}\int_{M^+} d^dp e^{ipx}S^{a-d}\Omega_{i\nu}(\omega\cdot\bar{e}) = {\Gamma({a-{d-2\over 2}+i\nu\over 2})\Gamma({a-{d-2\over 2}-i\nu\over 2})\over (-x^2)^{a\over 2}}\Omega_{i\nu}(e\cdot\bar{e}). 
}
Using this identity with $a=2\DL+J-1$ and the disconnected correlator in \TransfDisc, \afd\ gives 
\eqn\afdSecond{\eqalign{
	G(x)\Big|_{\Delta,J}= \lambda ic_{\Delta, J}2^{J-1}&e^{-i\pi(J-1)\over 2}(-x^2)^{-2\DL-J+1\over 2}\cr
  &\times\int^{\infty}_{-\infty} d\nu {\Gamma({2\DL+J-{d\over 2}+i\nu\over 2})\Gamma({2\DL+J-{d\over 2}-i\nu\over 2})\over \nu^2+(\Delta-{d\over 2})^2}\Omega_{i\nu}(e\cdot\bar{e}).
}}
The integrand in \afdSecond\ has simple poles at $\pm i\nu = \Delta-{d\over 2}$ coming from the denominator as well as poles due to the $\Gamma$-functions. The latter corresponds to the exchange of the double-trace operators $[\OL\OL]_{n,l}$; we will not consider these since by definition they do not contribute to the stress tensor sector. One can perform the integral in \afdSecond\ by deforming the contour in the lower half-plane where, in particular, one picks up the pole at $i\nu = \Delta-{d\over 2}$. This gives the following contribution to the correlator:
\eqn\reasds{\eqalign{
	G(x) \Big|_{\Delta,J}=(-x^2)^{-\DL}\lambda \,p[\Delta,J]{ic_{\Delta, J}\Pi_{\Delta-1,d-1}(e\cdot\bar{e})\over (e^{i\pi\over 2}\sqrt{-x^2})^{J-1}}+\ldots
}}
with the ellipses denoting double-trace operators which will not contribute to the stress tensor sector $\GG(z,\zbar)$ and we have defined 
\eqn\pp{
  p[\Delta,J]= 2^{J-1}(\DL)_{\Delta+J-d\over2}(\DL-{d-2\over 2})_{-\Delta+J+d-2\over 2}.
}

In particular, we see that by Fourier transforming a contribution in momentum space of the form \corrMom, i.e.\ the disconnected correlator times a Regge conformal block, one finds from \reasds\ the following contribution to the stress tensor sector of the correlator:
\eqn\posSpaceCont{\eqalign{
  \GG(\sigma,\rho)\Big|_{\Delta,J} = \lambda \,p[\Delta,J]\, g^{R}_{\Delta,J}(\sqrt{-x^2},e\cdot\bar{e}),
}}
valid to subleading order in $\sigma\to0$ and we have defined the position space Regge conformal block
\eqn\posSpaceReggeBlock{
  g^{R}_{\Delta,J}(\sqrt{-x^2},e\cdot\bar{e}) = ic_{\Delta, J}{\Pi_{\Delta-1,d-1}(e\cdot\bar{e})\over (e^{i\pi\over 2}\sqrt{-x^2})^{J-1}}.
}
Note that in \posSpaceCont, we have used the relation \gCylinder\ between the correlator on the cylinder and $\GG(z,\zbar)$ which is valid to subleading order in the Regge limit. 

In particular, we will be interested in $d=4$ where \posSpaceReggeBlock\ can be written in terms of $(z,\zbar)$ as 
\eqn\posSpaceReggeBlockZZbar{
  g^{R}_{\Delta,J}(z,\zbar) = i\bar{c}_{\Delta, J}e^{i\pi(1-J)\over 2}\Big({\log z\over\log\zbar}\Big)^{-(\Delta-1)\over 2}{(-\log z\log\zbar)^{1-J\over 2}\over 1-{\log \zbar\over\log z}},
}
which to subleading order in the Regge limit $\sigma\to0$ reduces to 
\eqn\posSpaceReggeBlockSigmaFourD{\eqalign{
  g^{R}_{\Delta,J}(\sigma,\rho) = i\bar{c}_{\Delta,J}{e^{-(\Delta-1)\rho}\over \sigma^{J-1}(1-e^{-2\rho})}\Big[&1-{\sigma\over 4}\Big((\Delta+J-2)e^{\rho}+(2+J-\Delta)e^{-\rho}\Big)\cr
  &+\OO(\sigma^2)\Big].
}}
Comparing the position space Regge conformal block in \posSpaceReggeBlockSigmaFourD\ with the conformal block in the Regge limit \reggeBlock, it is seen that in four dimensions, the former can to subleading order in $\sigma\to0$ be identified with a conformal block $g^{\circlearrowright}_{\Delta,J}(\sigma,\rho)$. To leading order this holds in any dimension, i.e., using the relation between $(\sigma,\rho)$ and $(z,\zbar)$ in \coordCorr\ and the known form of the conformal blocks \BlockGenerald, the contribution to the stress tensor sector $\GG(z,\zbar)$ in \posSpaceCont\ can be identified with\foot{We have not checked if this holds also at subleading order in arbitrary dimensions.}: 
\eqn\ResSingleTransform{
  \GG(\sigma,\rho)\Big|_{\Delta,J}= \lambda 2^{J-1}(\DL)_{\Delta+J-d\over2}(\DL-{d-2\over 2})_{-\Delta+J+d-2\over 2}g^{\circlearrowright}_{\Delta,J}(\sigma,\rho)+\ldots.
}

In what follows, we describe how to decompose products of Regge conformal blocks into sums of Regge conformal blocks. As we will see in Section 4, this is relevant when one considers the exponentiation of the phase shift which, when expanded into a series, will result in products of Regge conformal blocks. After having decomposed these products into sums of Regge conformal blocks, it is straightforward to use \posSpaceCont\ to find the contribution to the stress tensor sector of the correlator. We further note that while the phase-shift is only known to leading order in $S\gg1$, the leading and next-to-leading singularities in the Regge limit $\sigma\to0$ are not affected by subleading corrections to the phase shift. This is discussed further in Section 4.2 and 4.3.

\subsec{Two dimensions}
Consider a Regge conformal block in two dimensions:
\eqn\RCBTwoD{
	g^{\rm R}_{\Delta,J} = i\bar{c}_{\Delta,J} S^{J-1}e^{-(\Delta-1)L},
}
where $\bar{c}_{\Delta,J}$ a constant given in \cBarDef. A product of Regge conformal blocks with $(\Delta_i,J_i)$ weighted with constants $\lambda_i$ is trivially given by:
\eqn\prodRCB{
	\prod_{i=1}^{p} \lambda_i g^{\rm R}_{\Delta_i,J_i} = i^{p-1}\lambda g^{\rm R}_{\Delta,J}(S,L),
}
with 
\eqn\DataTwod{\eqalign{
	\Delta &= \sum_{i=1}^p \Delta_i-(p-1)\cr 
	J &= \sum_{i=1}^p J_i-(p-1)\cr 
	\lambda &= {1\over c_{\Delta,J}}\prod_{i=1}^p \lambda_i\bar{c}_{\Delta_i,J_i} .
}}
From \prodRCB, it is seen that the product of Regge conformal blocks in $d=2$ is also a Regge conformal block with $(\Delta,J,\lambda)$ given by \DataTwod. Assume a contribution in momentum space of the form
\eqn\resProdTwoD{
	{\BB(p)\over \BB_0(p)}\Big|_{\{\Delta_i,J_i\}}:= \prod_{i=1}^{p} \lambda_i g^{\rm R}_{\Delta_i,J_i}(S,L).
}
Using \prodRCB-\DataTwod, it follows from the Fourier transform in \posSpaceCont\ that the product of Regge conformal blocks in \resProdTwoD\ gives the following contribution to the stress tensor sector to subleading order in $\sigma$:
\eqn\ResProdTwoDPos{
  \GG(\sqrt{-x^2},e\cdot\bar{e})\Big|_{\{\Delta_i,J_i\}}=i^{p-1}\lambda 2^{J-1}(\DL)_{\Delta+J-2\over2}(\DL)_{-\Delta+J\over 2}g^{R}_{\Delta,J}(\sqrt{-x^2},e\cdot\bar{e}).
} 
Because a product of Regge conformal blocks in two dimensions is again a Regge conformal block, we see that it is trivial to perform the Fourier transform.

\subsec{Four dimensions}
In this section, products of Regge conformal blocks in four dimensions are considered. In particular, the decomposition of such products into a sum of Regge conformal blocks is described. Using this decomposition, one can do the Fourier transform using \posSpaceCont. 

A Regge conformal block in four dimensions is given by:
\eqn\RCBFourD{
	g^{\rm R}_{\Delta,J} = i\bar{c}_{\Delta,J}{S^{J-1}e^{-(\Delta-1)L}\over 1-e^{-2L}}.
}
Consider a product of $p$ Regge conformal blocks with scaling dimension and spin $(\Delta_i,J_i)$, $i=1,2,\ldots,p$, together with some weights $\lambda_i$: 
\eqn\ProdRCBFourD{
	\prod_{i=1}^{p} \lambda_i g^{\rm R}_{\Delta_i,J_i} = i^{p-1}{S^{J-1}e^{-(\Delta_0-1)L}\over (1-e^{-2L})^p}\prod_{i=1}^p \lambda_i\bar{c}_{\Delta_i,J_i},
}
where
\eqn\scalSpin{\eqalign{
  \Delta_0 &= \sum_{i=1}^p\Delta_i-(p-1),\cr
  J &= \sum_{i=1}^p J_i-(p-1).
}}
Expanding the factor $(1-e^{-2L})^{-p+1}$ in \ProdRCBFourD\ into a sum, the product of Regge conformal blocks in \ProdRCBFourD\ can be written as 
\eqn\ProdRCBFourDExp{
	\prod_{i=1}^{p} \lambda_i g^{\rm R}_{\Delta_i,J_i} = {i^{p-1}S^{J-1}e^{-(\Delta_0-1)L}\over (1-e^{-2L})}\prod_{i=1}^p \lambda_i\bar{c}_{\Delta_i,J_i}\sum_{n=0}^\infty {n+p-2\choose n}e^{-2nL}.
}
Compared to the two-dimensional case, it is seen from \ProdRCBFourD-\ProdRCBFourDExp\ that products of Regge conformal blocks in four dimensions decompose into an infinite sum of Regge conformal blocks with dimensions $\Delta_n=\Delta_0+2n$ and spin $J$. Explicitly, the product of Regge conformal blocks have the following decomposition: 
\eqn\prodOfBlocksFourD{
	\prod_{i=1}^{p} \lambda_i g^{\rm R}_{\Delta_i,J_i}(S,L) = i^{p-1}\sum_{n=0}^\infty \lambda_n g^{\rm R}_{\Delta_n,J}(S,L),
}
with 
\eqn\DataFourD{\eqalign{
	\Delta_n &= \sum_{i=1}^p\Delta_i+2n-(p-1),\cr
	J &= \sum_{i=1}^p J_i-(p-1),\cr
	\lambda_n &=  {1\over c_{\Delta_n,J}}{n+p-2\choose n}\prod_{i=1}^p \lambda_i\bar{c}_{\Delta_i,J_i}.
}}

Using the decomposition \prodOfBlocksFourD, it is straightforward to write down the Fourier transform of products of Regge conformal blocks using \posSpaceCont. Explicitly, a term in momentum space of the form \prodOfBlocksFourD\ 
\eqn\sad{\eqalign{
	{\BB(p)\over \BB_0(p)}\Big|_{\{\Delta_i,J_i\}} := \prod_{i=1}^{p} \lambda_i g^{\rm R}_{\Delta_i,J_i}(S,L),
}}
with $\Delta_n,J,\lambda_n$ given by \DataFourD, gives the following contribution to the stress tensor sector:
\eqn\sadPos{
  \GG(\sqrt{-x^2},e\cdot\bar{e})\Big|_{\{\Delta_i,J_i\}} =i^{p-1}\sum_{n=0}^\infty p[\Delta_n,J]\lambda_n g^{R}_{\Delta_n,J}(\sqrt{-x^2},e\cdot\bar{e}),
} 
to subleading order in $\sigma$. Here $p[\Delta,J]$ is the product of Pochhammer symbols defined in \pp.

\newsec{Regge limit of the stress tensor sector and the bulk phase shift}
\noindent In this section, the heavy-heavy-light-light correlator is studied assuming that the correlator in momentum space is given by
\eqn\assumpExp{
  \BB(p)= \BB_0(p)e^{i\delta(S,L;\mu)},
}
where $\delta(S,L;\mu)$ is the bulk phase shift. The phase shift was calculated to all orders in a perturbative expansion in $\mu$ in \KulaxiziDXO:
\eqn\phaseShift{
  \delta(S,L;\mu)=\sum_{k=0}^{\infty}\mu^k\delta^{(k)}(S,L).
}
It was further shown in \KarlssonQFI\ that $\delta^{(k)}:=\sum_{n=0}^\infty\delta^{(k)}_n$ can be decomposed in terms of Regge conformal blocks as
\eqn\PhaseShiftRCB{\eqalign{
i\,\delta^{(k)}_n(S,L)&=  \,\, f(k)\,\,\lambda_k(n) \,\, g_{\scriptscriptstyle \tau_0(k)+2n+2,2}^{R}(S,L)  \cr
\lambda_k(n)&=a(n) \, {2^{-4 n}\left[\left({\tau_0(k)+4\over 2}\right)_n\right]^2\over \left({\tau_0(k)+3\over 2}\right)_n\left({\tau_0(k)+5\over 2}\right)_n} ,\quad \tau_0(k)=k(d-2)\,
}}
with
\eqn\defak{\eqalign{f(k)&= {\sqrt{\pi}\over 64} \,\, {1\over 2^{k(d-2)} \,\,k!}  \, {\Gamma\left({k d+1\over 2}\right)\Gamma\left({k(d-2)+4\over 2}\right)\over\Gamma\left({k(d-2)+5\over 2}\right)\Gamma\left({k(d-2)+3\over 2}\right)}\,,\cr
  a(n)&= {2^{2n}\over n!}{\tau_0(k)+2\over \tau_0(k)+2+2n}{({\tau_0(k)-d+2\over 2})_{n}({\tau_0(k)+1\over 2})_n\over (\tau_0(k)+n+2-{d\over 2})_n}. 
}}
Note that $\lambda_1(n)=0$ for $n=1,2,\ldots$ implying that the first-order phase shift reduces to a single term in \PhaseShiftRCB. Expanding the exponential in \assumpExp\ results in a sum of products of Regge conformal blocks. Using the decomposition of such products in four dimensions described in Section 3, we read off the contribution to the stress tensor sector $\GG$ of the correlator from the phase shift. 

At $k$-th order, the stress tensor sector of the correlator behaves as
\eqn\RegLim{
  \GG^{(k)}(\sigma,\rho) = {F_{k,{\rm L}}(\rho)\over \sigma^k}+{F_{k,{\rm NL}}(\rho)\over\sigma^{k-1}}+\OO(\sigma^{-k+2})\qquad \quad \sigma\to 0,\quad \rho{\rm-fixed},
}
for some functions $F_{k,{\rm L}}(\rho)$ and $F_{k,{\rm NL}}(\rho)$ in the Regge limit. The leading and next-to-leading Regge singularity of the stress tensor sector of the correlator $\GG^{(k)}$ at $\OO(\mu^k)$ were defined in \defReggeSing\ by
\eqn\defReggeSing{\eqalign{
  &{\rm Leading\,\,Regge\,\,singularity:} \qquad {F_{k,{\rm L}}(\rho)\over \sigma^k},\cr
  &{\rm Next{\rm-}to{\rm-}leading\,\,Regge\,\,singularity:} \qquad {F_{k,{\rm NL}}(\rho)\over \sigma^{k-1}}.
}}
By expanding \assumpExp\ and Fourier transforming terms propoportional to $S^k$ and $S^{k-1}$ at $\OO(\mu^k)$, the leading- and next-to-leading singularities are found to all orders in $\mu$ at finite impact parameter. In particular, the leading singularities in the Regge limit comes from the exponentiation of the first-order phase shift. We find perfect agreement with the calculation of a light particle propagating in a shockwave background in \FitzpatrickEFK. It is then shown, from the exchange of stress tensor, that there is no correction to $\delta^{(1)}$ of $\OO(S^0)$ for large $S\gg1$. Using this knowledge, we calculate the next-to-leading Regge singularities to all orders in $\mu$ at finite impact parameter. Both the leading and next-to-leading order Regge singularities agree in the large impact parameter limit with known results obtained using lightcone bootstrap \refs{\KulaxiziTKD,\KarlssonDBD}.

\subsec{Leading Regge singularities}
In this section, the leading terms in the correlator as $\sigma\to0$, which were defined in \defReggeSing\ as the leading Regge singularities, at each order in $\mu$ are studied in four dimensions. Expanding \assumpExp, these come from the exponentiation of the first-order phase shift $\delta^{(1)}$:
\eqn\expFirstOrder{\eqalign{
  \BB(p)&=\BB_0(p) e^{i\mu\delta^{(1)}}+\ldots\cr
        &= \BB_0(p)\sum_{k=0}^{\infty}\mu^k\Big[{i^k\over k!}(\delta^{(1)})^k+\OO(S^{k-1})\Big].
}}
A term proportional to $S^k$ will, after Fourier transform to position space, scale as $\sigma^{-k}$ when $\sigma\to0$. This will be the leading Regge singularity at $\OO(\mu^k)$.

The first-order phase shift is given by \PhaseShiftRCB\
\eqn\phaseShiftFirs{\eqalign{
  i\delta^{(1)} &={1\over 240}g_{4,2}(S,L)\cr
               &=i{3\pi\over 2}{S e^{-3L}\over 1-e^{-2L}}.
}}
The term at $\OO(\mu^k)$ in \expFirstOrder\ is a product of $k$ Regge conformal blocks with dimension $\Delta=4$ and spin $J=2$. Using the decomposition of products of Regge conformal blocks \prodOfBlocksFourD-\DataFourD, the expansion of the momentum space correlator in \expFirstOrder\ can be written in terms of Regge conformal blocks with
\eqn\dataFirstOrder{\eqalign{
  \Delta_{k,n}&=3k+2n+1,\cr
  J_k &= k+1,
}}
where $n=0,1,\ldots$. 
Using \posSpaceCont\ to do the Fourier transform of each Regge conformal block, this gives the following contribution to the stress tensor sector of the correlator: 
\eqn\leadingReggeFull{\eqalign{
  \GG^{(k)}(\sigma,\rho)\Big|_{(\delta^{(1)})^k}=\sum_{n=0}^\infty{1\over k!}&\left({i\bar{c}_{4,2}\over 240}\right)^{k}{n+k-2\choose n}(\DL)_{\Delta_{k,n}+J_k-4\over 2}(\DL-1)_{-\Delta_{k,n}+J_k+2\over 2}\cr
  \times &{(-i)\over c_{\Delta_{k,n},J_k}}g^R_{\Delta_{k,n},J_k}(\sqrt{-x^2},e\cdot\bar{e}),
}}
valid to subleading order in $\sigma$ with $k=1,2,\ldots$.

The leading Regge singularities can be written in terms of $(\sigma,\rho)$ using $\sqrt{-x^2}\approx-i\sigma$ and $-e\cdot\bar{e} \approx\cosh\rho$. From \leadingReggeFull\ one finds:
\eqn\FirstPH{\eqalign{
  \GG^{(k)}(\sigma,\rho)\Big|_{(\delta^{(1)})^k}= \sum_{n=0}^{\infty}&{(3i\pi)^k\over k!}{n+k-2\choose n}(\DL)_{\Delta_{k,n}+J_k-4\over 2}(\DL-1)_{-\Delta_{k,n}+J_k+2\over 2}\cr
  &\times{e^{-(\Delta_{k,n}-1)\rho}\over \sigma^{J_k-1}(1-e^{-2\rho})}+\ldots,
}}
where the ellipses denote terms subleading in $\sigma\to0$. Explicitly, inserting the dimensions and spins $(\Delta_{k,n},J_k)$ given in \dataFirstOrder, we find
\eqn\FirstPhRewritten{\eqalign{
  \GG^{(k)}(\sigma,\rho)\Big|_{(\delta^{(1)})^k}=& \sum_{n=0}^{\infty}{(3i\pi)^k\over k!}{n+k-2\choose n}(\DL)_{2k+n-1}(\DL-1)_{1-k-n}{e^{-(3k+2n)\rho}\over \sigma^{k}(1-e^{-2\rho})}\cr
  &+\ldots.
}}
The sum over $n$ can further be written as a hypergeometric function:
\eqn\FirstPhSecRewr{\eqalign{
  \GG^{(k)}(\sigma,\rho)\Big|_{(\delta^{(1)})^k}=&{(3i\pi)^k\over k!\sigma^k(1-e^{-2\rho})}(\DL)_{2k-1}(\DL-1)_{1-k}\cr
    &\times e^{-3k\rho}{}_2F_1(k-1,\DL+2k-1;-\DL+k+1;-e^{-2\rho})+\ldots.
}}
These are the leading Regge singularities, i.e., terms that behave as $\sigma^{-k}$ at $\OO(\mu^k)$, to all orders in $\mu$ at finite impact parameter. The result \FirstPhSecRewr\ agrees with the calculation in a shockwave background in \FitzpatrickEFK, for details, see Appendix A. In particular, consider the terms in the sum in \FirstPhSecRewr\ with $k=2,3$:
\eqn\leadRegToThirdOrder{\eqalign{
  \GG^{(2)}(\sigma,\rho)\Big|_{\sigma^{-2}} &\rhoequal -{9\pi^2\DL(\DL+1)(\DL+2)\over 2(\DL-2)}{e^{-6\rho}\over \sigma^2},\cr
  \GG^{(3)}(\sigma,\rho)\Big|_{\sigma^{-3}} &\rhoequal -{9i\pi^3\DL(\DL+1)(\DL+2)(\DL+3)(\DL+4)\over 2(\DL-2)(\DL-3)}{e^{-9\rho}\over \sigma^3},
}}
where we have further taken the large impact parameter limit $\rho\to\infty$. The leading Regge singularities in \leadRegToThirdOrder\ agree with those in \ratioCorrSecond-\ratioCorrThird; the latter were found using lightcone bootstrap \refs{\KulaxiziTKD,\KarlssonDBD} and are due to minimal-twist double-stress and triple-stress tensors.

We note that the first-order phase shift is to leading order in $\sigma$ fixed by the exchange of stress tensor in the direct channel in the CFT \KulaxiziDXO. It is therefore universally fixed by Ward identities and does not depend on higher derivative corrections to the gravity action. This is discussed further in Section 4.2.

It is seen that the leading Regge singularities in \FirstPhRewritten, which can be identified with the leading behaviour of a conformal block in the Regge limit with dimension $\Delta_{k,n}=3k+2n+1$ and spin $J_k=k+1$, have poles and zeroes specified by the Pochhammer symbols to be given by:
\eqn\polesAndZeroes{\eqalign{
  {\rm Zeroes:}\qquad \DL &= -(2k+n-2),-(2k+n-3),\ldots,0\cr
  {\rm Poles:}\qquad \DL  &= 2,3,\ldots,k+n.
}}
The position of the poles and zeroes are seen to be related to the dimension and spin of the blocks that are present in the decomposition of $(\delta^{(1)})^k$. Possible implications of the position of poles and zeroes were discussed in \FitzpatrickEFK. In particular, it is expected that the OPE coefficients of multi-stress tensors with minimal-twist have the same poles as predicted by \polesAndZeroes\ with $n=0$. This agrees with the results in \refs{\FitzpatrickZQZ,\KulaxiziTKD,\KarlssonDBD}. Moreover, we further expect from \polesAndZeroes\ the OPE coefficients for non-minimal-twist multi-stress tensors to have poles at $\Delta=2,3,\ldots,k+n$, with $n$ being related to the twist by $\tau=k(d-2)+2n$.

\subsec{The first-order phase shift and the stress tensor exchange}
The phase shift in \phaseShift\ calculated in the bulk is linear in the energy $S\gg1$. In principle, it could receive corrections in an ${1\over S}$ expansion that will be important when expanding \assumpExp. On the other hand, from the CFT point of view, the stress tensor is the only operator that appears at $\OO(\mu)$ in the stress tensor sector. Using this, we show that there is no correction to $\delta^{(1)}$ in four dimensions of order $\OO(S^0)$. This will important in Section 4.3 where the next-to-leading Regge singularities are studied.

The stress tensor exchange in four dimensions is found using the knwon OPE coefficients in \opeStress\ and the conformal block given in \TChannelBlock. Explicitly, one finds the following contribution as $\sigma\to0$: 
\eqn\ReggeStressTensor{\eqalign{
  P^{(HH,LL)}_{T_{\mu\nu}}g^{\circlearrowright}_{4,2}(\sigma,\rho) = \mu {3\pi i \DL e^{-3\rho}\over(1-e^{-2\rho})}{1\over\sigma}-\mu {3\pi i \DL e^{-2\rho}\over(1-e^{-2\rho})}+\OO(\sigma).
}}

On the other hand, expanding the momentum space correlator in \assumpExp\ one finds at $\OO(\mu)$:
\eqn\ordermu{
  \BB(p)=\BB_0(p)i\mu\delta^{(1)}+\OO(\mu^2),
} 
with the first-order phase shift in $d=4$ given in \phaseShiftFirs. Fourier transforming \ordermu\ using \posSpaceCont\ gives the following contribution to the correlator in position space:
\eqn\deltaOneContSigma{\eqalign{
  \GG(\sigma,\rho)\Big|_{\delta^{(1)}}=\mu {3\pi i \DL e^{-3\rho}\over(1-e^{-2\rho})}{1\over\sigma}-\mu{3\pi i \DL e^{-2\rho}\over(1-e^{-2\rho})}+\OO(\sigma).
}}
where we used $p[4,2]=2\DL$. Comparing the contribution from the stress tensor in the Regge limit \ReggeStressTensor, with the contribution from $\delta^{(1)}$ in \deltaOneContSigma, we find that both the leading and next-to-leading terms as $\sigma\to0$ agree\foot{Since the leading terms were known to agree, this follows immediately from the observation below \posSpaceReggeBlockSigmaFourD.}. This shows that there is no $\OO(S^0)$ correction to the first-order phase shift. 

\subsec{Next-to-leading Regge singularities}
In this section, the next-to-leading Regge singularities are considered, i.e.\ terms proportional to $\sigma^{1-k}$ at $\OO(\mu^k)$, to all orders in $\mu$. These will be due to terms in \assumpExp\ of the form $(\delta^{(1)})^k$ that were calculated in Section 4.2, and terms of the form $(\delta^{(1)})^{k-2}\delta^{(2)}$ which are of $\OO(S^{k-1})$ for $S\gg 1$. Note that it was shown in Section 4.2 that there is no $\OO(S^0)$ correction to the first-order phase shift. The contribution to the next-to-leading Regge singularities from terms of the form $(\delta^{(1)})^k$ are therefore given by \leadingReggeFull.

Consider terms in \assumpExp\ of the form $(\delta^{(1)})^{k-2}\delta^{(2)}$:
\eqn\subterms{
  {\BB(p)\over \BB_0(p)}\Big|_{\delta^{(2)}(\delta^{(1)})^{k-2}} = {\mu^ki^{k-1}\over (k-2)!}(\delta^{(1)})^{k-2}\delta^{(2)},
}
with $k=2,3,\ldots$.
Inserting the decomposition of $\delta^{(2)}$ from \PhaseShiftRCB\
\eqn\deltaTwo{
  i\delta^{(2)}=f(2)\sum_{n=0}^\infty \lambda_2(n)g_{4+2n,2}^{R}(S,L)
}
and the first-order phase shift \phaseShiftFirs, we rewrite \subterms\ as
\eqn\subtermsDecomp{
   {\BB(p)\over \BB_0(p)}\Big|_{\delta^{(2)}(\delta^{(1)})^{k-2}} = {\mu^ki^{k-1}\over (k-2)!}\Big({3\pi\over 2}\Big)^{k-2}\sum_{n=0}^\infty f(2)\lambda_2(n)\bar{c}_{6+2n,2}{S^{k-1}e^{-(3k+2n-1)L}\over (1-e^{-2L})^{k-1}}.
}
Expanding $(1-e^{-2L})^{-k+2}$, we find
\eqn\subTermsSecondDecomp{\eqalign{
  {\BB(p)\over \BB_0(p)}\Big|_{\delta^{(2)}(\delta^{(1)})^{k-2}} =f(2){\mu^ki^{k-1}\over (k-2)!}\Big({3\pi\over 2}\Big)^{k-2}\sum_{n,m=0}^\infty &{m+k-3\choose m}\lambda_2(n)\bar{c}_{6+2n,2}\cr
  &\times{S^{k-1}e^{-(\Delta_{n,m}-1)L} \over 1-e^{-2L}},
}}
with 
\eqn\DeltaNM{
  \Delta_{n,m}=3k+2(n+m).
}
Comparing the product of Regge conformal blocks in \subterms\ with \subTermsSecondDecomp, it is seen that the latter is a decomposition into Regge conformal blocks with dimensions $3k+2(n+m)$ and spin $k$. This can conveniently be organized into blocks with different twists
\eqn\relabSum{\eqalign{
  {\BB(p)\over \BB_0(p)}\Big|_{\delta^{(2)}(\delta^{(1)})^{k-2}} =f(2){\mu^ki^{k-1}\over (k-2)!}\Big({3\pi\over 2}\Big)^{k-2}\sum_{p=0}^\infty\sum_{n=0}^p &{k+p-n-3\choose p-n}\lambda_2(n)\bar{c}_{6+2n,2}\cr
  &\times{S^{k-1}e^{-(3k+2p-1)L}\over 1-e^{-2L}}.
}} 

To get the next-to-leading order Regge singularities from \relabSum, it is enough to use the leading order relation $\sqrt{-x^2}=-i\sigma$ and $-e\cdot\bar{e}=\cosh\rho$. This is so since terms in \relabSum\ are of $\OO(S^{k-1})$ and therefore start to contribute at $\sigma^{-k+1}$ in position space. Using \posSpaceCont\ to perform the Fourier transform of each term in the sum, one finds that \relabSum\ gives the following contribution to the next-to-leading order Regge singularities in the stress tensor sector:
\eqn\resSub{\eqalign{
  \GG^{(k)}(\sigma,\rho)|_{\delta^{(2)}(\delta^{(1)})^{k-2}} =2f(2)&\sum_{p=0}^\infty\sum_{n=0}^p {i^{k-1}(3\pi)^{k-2}\over (k-2)!}(\DL)_{2k+p-2}(\DL-1)_{1-k-p}\cr
  \times& {k+p-n-3\choose p-n}\lambda_2(n)\bar{c}_{6+2n,2}{e^{-(3k+2p-1)\rho}\over \sigma^{k-1}(1-e^{-2\rho})}\cr
  &+\ldots,
}}
where the ellipses denote subleading corrections in $\sigma$. To get the full result for the next-to-leading Regge singularities we need to add the contribution from \leadingReggeFull. This is found using the correction to the position Regge conformal block \posSpaceReggeBlockSigmaFourD\ and the leading order expression \FirstPhRewritten
\eqn\subCorrNext{\eqalign{
\GG(\sigma,\rho)^{(k)}\Big|_{(\delta^{(1)})^k,\sigma^{-k+1}}=&{1\over2}\sum_{n=0}^{\infty}{(3i\pi)^k\over k!}{n+k-2\choose n}(\DL)_{2k+n-1}(\DL-1)_{1-k-n}\cr
&\times\Big[(k+n-1)e^{-\rho}-(2k+n)e^\rho\Big]{e^{-(3k+2n)\rho}\over \sigma^{k-1}(1-e^{-2\rho})}.
}}
The next-to-leading Regge singularities to all orders in $\mu$ at finite impact parameter is therefore given by the sum of \resSub\ and \subCorrNext.  

Consider the large impact parameter limit in which only the $p=n=0$ term in \resSub\ contributes. In this limit, \resSub\ reduces to
\eqn\resSubleadingLargeRho{\eqalign{
  \GG(\sigma,\rho)^{(k)}|_{\delta^{(2)}(\delta^{(1)})^{k-2}} \rhoequal{35\over 6}{(3i\pi)^{k-1}\over (k-2)!}(\DL)_{2k-2}(\DL-1)_{1-k}{e^{-(3k-1)\rho}\over \sigma^{k-1}}.
}}
This is the contribution of $\delta^{(2)}(\delta^{(1)})^{k-2}$ to the next-to-leading Regge singularity at $k$-th order in the large impact parameter limit. 

Including the contribution to the next-to-leading Regge singularity from \subCorrNext\ due to $(\delta^{(1)})^k$ together with \resSubleadingLargeRho, we find for large impact parameter at $\OO(\mu^2)$:
\eqn\nextSecondOrder{
  \GG^{(2)}(\sigma,\rho)\Big|_{\sigma^{-1}} = \Big[{35i\pi\DL(\DL+1)\over 2(\DL-2)}+{18\pi^2\DL(\DL+1)(\DL+2)\over 2(\DL-2)}\Big]{e^{-5\rho}\over \sigma}.
}
Likewise, consider the next-to-leading singularity at $\OO(\mu^3)$ which using \subCorrNext\ and \resSubleadingLargeRho\ gives
\eqn\nextThirdOrder{\eqalign{
  \GG^{(3)}(\sigma,\rho)\Big|_{\sigma^{-2}} = \Big[&-{105\pi^2\DL(\DL+1)(\DL+2)(\DL+3)\over 2(\DL-2)(\DL-3)}\cr
  &+{27i\pi^3\DL(\DL+1)(\DL+2)(\DL+3)(\DL+4)\over 2(\DL-2)(\DL-3)}\Big]{e^{-8\rho}\over\sigma^2}.
}}
Comparing the next-to-leading Regge singularities in the large impact parameter limit, \nextSecondOrder\ and \nextThirdOrder, with \ratioCorrSecond-\ratioCorrThird, respectively, we find agreement between the result obtained here using the phase shift and known results obtained using lightcone bootstrap. 

Similarily to the leading Regge singularities, the next-to-leading singularities due to $\delta^{(2)}(\delta^{(1)})^{k-2}$ have a simple dependence on the scaling dimension $\DL$ -- the poles and zeroes are fixed by the dimension and spin of the Regge conformal blocks appearing in the decomposition \relabSum. From \resSub, the poles and zeroes are found to be given by:
\eqn\polesAndZeroesSub{\eqalign{
  {\rm Zeroes:}\qquad \DL &= -(2k+p-3),-(2k+p-3),\ldots,0\cr
  {\rm Poles:}\qquad \DL  &= 2,3,\ldots,k+p.
}}
Note that the poles are the same as those for the leading Regge singularities in \polesAndZeroes. 

\newsec{Discussion}
Using the first- and second-order phase shift, we derived the leading and next-to-leading Regge singularities of the stress tensor sector to all orders. The leading Regge singularity at each order was shown to be determined by the first-order phase shift. This is universally fixed by the stress tensor exchange and our results agree with the expression obtained in \FitzpatrickEFK. The next-to-leading Regge singularity at each order further depends on the second-order phase shift. In general, the second-order phase shift is expected to be non-universal in the sense that it depends on higher derivative corrections to the gravitational action in the bulk.

It has been argued in \refs{\FitzpatrickZQZ,\FitzpatrickEFK,\HuangFOG,\KarlssonDBD} that the minimal-twist multi-stress tensor sector of CFTs with large central charge is universal. This was argued from the holographic point of view in \refs{\FitzpatrickZQZ,\FitzpatrickEFK}. There the two-point function of a minimally coupled scalar propagating in an AdS black hole background was studied in higher derivative gravity. In \KarlssonDBD\ it was shown that the ansatz \gkhd\ solves the crossing relations and that the minimal-twist subsector of the stress tensor sector is, therefore, determined in terms of the exchange of the stress tensor. Since the stress tensor exchange is fixed by Ward identities, this implies that the minimal-twist subsector is universal. In terms of the phase shift, this would imply that when decomposing the phase shift in terms of Regge conformal blocks, the contribution proportional to the block with the lowest twist at each order is universal. It would be interesting to study explicitly the effect of higher derivative terms on the phase shift and verify this. Universality in the minimal-twist sector would imply that the large impact parameter limit of our results for the next-to-leading Regge singularities is universal.

While we have focused on $d=4$, it would be interesting to understand how to extend this to general dimensions. In particular, in $d=6$, the hyperbolic space propagators take a similar form as in $d=4$ and it would be interesting to find a similar decomposition of products of Regge conformal blocks. Moreover, for large impact parameter $L$, the hypergeometric function in \HyperbolicPropagator\ can be set to $1$ in any dimension. In this limit the Regge conformal blocks in any dimension resemble the two-dimensional blocks. 

Consider the exponentiation of the phase shift in $d=4$ at $\OO(\mu^3)$:
\eqn\expFull{\eqalign{
  \BB(p)\Big|_{\mu^3}=\BB_0(p)\Big[-i{(\delta^{(1)})^3\over 3!}-\delta^{(1)}\delta^{(2)}+i\delta^{(3)}\Big].
}}
The leading and next-to-leading Regge singularities obtained from \expFull\ were already discussed in Section 4.1 and 4.3 respectively. In Appendix B, it is shown that including the first subleading correction for $\sigma\to0$ to the Fourier transform of the $\delta^{(1)}\delta^{(2)}$ term in \expFull, one finds for $\rho\to\infty$
\eqn\subleadingNextNext{
  \GG^{(3)}(\sigma,\rho)\Big|_{\delta^{(1)}\delta^{(2)},\sigma^{-1}}\rhoequal {525\pi^2\DL(\DL+1)(\DL+2)(\DL+3)\over 4(\DL-2)(\DL-3)}{e^{-7\rho}\over \sigma}.
}
This agrees with the third line in \ratioCorrThirdsigmaOne\ obtained using lightcone bootstrap. More interesting is the last term in \expFull\ given by the phase shift at third order. It gives the following contribution to the stress tensor sector of the correlator:
\eqn\thirdOrderPhaseShift{
  \GG^{(3)}(\sigma,\rho)\Big|_{\delta^{(3)}}=f(3)\sum_{n=0}^\infty \lambda_3(n) p[\tau_0(3)+2n+2,2]\,\, g_{\scriptscriptstyle \tau_0(3)+2n+2,2}^{\circlearrowright}(\sigma,\rho),
}
to leading order in $\sigma\to0$ and $\lambda_{k}(n)$ is given by the decomposition of the phase shift in \PhaseShiftRCB. In particular, in the large impact parameter limit only the $n=0$ term contributes:
\eqn\deltaThree{
  \GG^{(3)}(\sigma,\rho)\Big|_{\delta^{(3)}}\rhoequal {1155i\pi\DL(\DL+1)(\DL+2)\over 8(\DL-2)(\DL-3)}{e^{-7\rho}\over \sigma}.
}
This agrees with the term in the first line in \ratioCorrThirdsigmaOne\ due to minimal-twist triple-stress tensors obtained from lightcone bootstrap. The remaining term in \ratioCorrThirdsigmaOne\ presumably comes from subsubleading corrections to $(\delta^{(1)})^3$ as well as possible subleading corrections to the second-order phase shift.

Following the discussion\foot{A similar discussion was previously considered in \KarlssonQFI\ and we elaborate on it here.} above on the term linear in $S$ at $\OO(\mu^3)$, it is interesting to study terms linear in $S$ at any order in $\mu$:
\eqn\linearSAll{
	\BB(p)\Big|_{\mu^k,S} = \BB_0(p)i\delta^{(k)}.
}
The corresponding contribution to the stress tensor sector to leading order in $\sigma\to0$ can, in any dimension, be identified with the leading Regge behaviour of operators $\OO_{\Delta_{k,n},J=2}$ with scaling dimension and spin given by
\eqn\DeltaEff{\eqalign{
  &\Delta_{k,n}=k(d-2)+2n+2,\cr
  &J=2.
}}
We refer to these operators as effective in the sense that they are not necessarily present in the spectrum, but rather are due to the resummation of multi-stress tensor with arbitrary spin. The contribution linear in $S$ in \linearSAll\ is easily Fourier transformed using \posSpaceCont\ and the decomposition of the phase shift in terms of Regge conformal blocks \PhaseShiftRCB. Explicitly, it is found that \linearSAll\ gives the following contribution to the stress tensor sector of the correlator to leading order in $\sigma\to 0$ in any dimension $d$:
\eqn\linearSTransf{
\mu^k\GG^{(k)}(\sigma,\rho)\Big|_{\delta^{(k)}} = \mu^kf(k)\,\,\sum_{n=0}^\infty p[k(d-2)+2n+2,2]\lambda_k(n) \,\, g_{k(d-2)+2n+2,2}^{\circlearrowright}(\sigma,\rho),
}
where $\lambda_{k}(n)$ and $f(k)$ are given in \PhaseShiftRCB, $p[\Delta,J]$ is a combination of Pochhammer symbols defined in \pp\ and $g_{\Delta,J}^{\circlearrowright}(\sigma,\rho)$ is the leading contribution of a conformal block in the Regge limit. Interpreting each term in \linearSTransf\ as due to the exchange of an effective operator $\OO_{\Delta_{k,n},2}$, the coefficients in \linearSTransf\ are products of the corresponding OPE coefficients for such exchanges $P^{HH,LL}_{\OO_{\Delta_{k,n},J=2}}=\mu^kf(k)p[k(d-2)+2n+2,2]\lambda_k(n)$:
\eqn\OPEEffective{\eqalign{
  P^{HH,LL}_{\OO_{\Delta_{k,n},2}} = &\mu^k{\sqrt{\pi}(k(d-2)+2)({(k-1)(d-2)\over2})_n \Big[({k(d-2)+4\over 2})_n\Big]^2({k(d-2)+1\over 2})_n\over 2^{5+k(d-2)+2n}k!n!(k(d-2)+2n+2)({2k(d-2)-d+2n+4\over 2})_n}\cr
  &\times {\Gamma({k(d-2)+4\over 2})\Gamma({dk+1\over 2})\Gamma(\DL+{k(d-2)+2n-d+4\over 2})\Gamma(\DL-{k(d-2)+2n\over 2})\over \Gamma(\DL)\Gamma(\DL-{d-2\over2})\Gamma({k(d-2)+2n+3\over 2})\Gamma({k(d-2)+2n+5\over 2})}.
}}
In \KravchukHTV\ it was shown using conformal Regge theory that when the correlator is dominated by an isolated pole in the $J$-plane, the corresponding exchange is due to a light-ray operator. It would be interesting to understand if there is an interpretation of the operators $\OO_{\Delta_{k,n},2}$ mentioned here, which are directly related to the phase shift, in terms of such light-ray operators. See also \refs{\KologluBCO,\KologluMFZ}. Note that $\lambda_{k}(n)$ from \PhaseShiftRCB\ are valid assuming Einstein gravity in the bulk. While expected to be non-universal for general $n$, we expect the $\lambda_k(0)$ coefficient in the phase shift to be universal and therefore \OPEEffective\ with $n=0$ to be universal.

\bigskip
\bigskip

\noindent {\bf Acknowledgments}: 
I would like to thank M. Kulaxizi, A. Parnachev and P. Tadi\' c for discussions, collaboration on related work and comments on the manuscript. I further thank G-S. Ng for discussions and R. Pereira for comments on the manuscript. This work is supported in part by the Laureate Award IRCLA/2017/82 from the Irish Research Council. 

\appendix{A}{Comparing leading Regge singularities with the shockwave calculation}
In order to compare the leading Regge singularities in \FirstPhRewritten\ with the stress tensor sector calculated in \FitzpatrickEFK, the following identity is useful 
\eqn\identityShockwave{\eqalign{
  &{e^{2(\DL+{k\over 2})\rho}\over e^{2\rho}-1}{\Gamma(1-k)\Gamma(\DL+2k-1)\Gamma(2\DL+k)\over k!\Gamma(\DL)\Gamma(\DL-1)\Gamma(\DL+k)(2\DL+k-1)}\cr
  \times&\Bigg[\tilde{F}_{\DL,n,-1}+{(e^{2\rho}-1)(\DL+2k-1)\over \DL+k}\tilde{F}_{\DL,n,0}-{e^{2\rho}(\DL+2k-1)(\DL+2k)\over (\DL+k)(\DL+k+1)}\tilde{F}_{\DL,n,1}\Bigg]\cr
    = &{e^{-3k\rho}\over 1-e^{-2\rho}}{\Gamma(\DL-k)\Gamma(\DL+2k-1)\over k!\Gamma(\DL-1)\Gamma(\DL)}{}_2F_1(k-1,\DL+2k-1;-\DL+k+1,-e^{-2\rho}),
}}
where\foot{$\tilde{F}_{\DL,n,a}$ is related to $F_{\DL,n,a}$ in \FitzpatrickEFK\ if one uses their identity Eq.\ (44) and keep only the part relevant to the stress tensor sector and set $(\eta)_{\rm there}=(e^{-2\rho})_{\rm here}$.}
\eqn\FTilde{\eqalign{
  \tilde{F}_{\DL,n,a}(e^{-2\rho}) = &{\Gamma(\Delta-k-a)\Gamma(\DL+k+a+1)\over \Gamma(1-k)\Gamma(2\DL+k)}\cr
  \times &e^{-2(\DL+2k+a)\rho}{}_2F_1(\DL+2k+a,k;-\DL+k+a+1;-e^{-2\rho}).
}}
With \identityShockwave\ one can check that \FirstPhRewritten\ agrees with the contribution from the stress tensor sector for fixed $\rho$, or $\eta$, in \FitzpatrickEFK. 

\appendix{B}{Further comparison with lightcone results}
In this section, we further compare predictions obtained using the phase shift with known results in the lightcone limit. 

\subsec{Triple-stress tensors in four dimensions}
Consider the momentum space correlator \assumpExp\ at $\OO(\mu^3)$. In the large impact parameter limit, this is compared with the explicit resummation of minimal-twist triple-stress tensors discussed in Section 2. 

Consider the correlator \assumpExp\ at $\OO(\mu^3)$:
\eqn\corrMuCub{\eqalign{
  {\BB(p)\over \BB_0(p)}\Big|_{\mu^3}= -i{(\delta^{(1)})^3\over 3!}-\delta^{(1)}\delta^{(2)}+i\delta^{(3)}.
}}
The leading and next-to-leading singularities are due to the first two terms in \corrMuCub\ and were discussed in Section 4. At $\OO({1\over\sigma})$ there will be a contribution from the last term $i\delta^{(3)}$ in \corrMuCub. Using the decomposition of the phase shift in \PhaseShiftRCB, it is straightforward to use \posSpaceCont\ to find the corresponding contribution to the stress tensor sector:
\eqn\linearS{\eqalign{
  \GG^{(3)}(\sigma,\rho)|_{\delta^{(3)}} &= f(3)\sum_{n=0}^\infty \lambda_3(n) p[\tau_0(3)+2n+2,2]\,\, g_{\scriptscriptstyle \tau_0(3)+2n+2,2}^{\circlearrowright}(\sigma,\rho)+\ldots,
}}
in any dimension $d$ and the ellipses denote subleading corrections in $\sigma\to0$. Here $\tau_0(k)=k(d-2)$ is the minimal-twist of multi-stress tensors at $k$-th order.
To compare the large impact parameter limit with the contribution from minimal-twist multi-stress tensors, consider the term in \linearS\ with $n=0$:
\eqn\matchFourDLinear{
  \GG^{(3)}(\sigma,\rho)|_{\delta^{(3)}} \rhoequal{1155i\pi\DL(\DL+1)(\DL+2)\over 8(\DL-2)(\DL-3)}{e^{-7\rho}\over \sigma}.
}
We thus see that \matchFourDLinear\ agree with the first line in \ratioCorrThirdsigmaOne\ at $\OO(\mu^3)$ due to minimal-twist triple-stress tensors in $d=4$. 

There will also be a contribution at $\OO({1\over \sigma})$ due to the first subleading correction to the second term $-\delta^{(1)}\delta^{(2)}$ in \corrMuCub. It is straightforward to include the correction to the position space Regge conformal block in \posSpaceReggeBlockSigmaFourD\ to the expression \resSub\ found in Section 4.3. Taking the large impact parameter with $k=3$ one finds:
\eqn\nextCorrTriple{
  \GG^{(3)}(\sigma,\rho)|_{\delta^{(1)}\delta^{(2)},\sigma^{-1}} \rhoequal {525\pi^2\DL(\DL+1)(\DL+2)(\DL+3)\over 4(\DL-2)(\DL-3)}{e^{-7\rho}\over \sigma},
} 
which agree with the last line in \ratioCorrThirdsigmaOne\ obtained using lightcone bootstrap.

\subsec{Double-stress tensors in six dimensions}
Consider the correlator \assumpExp\ at $\OO(\mu^2)$ in $d=6$:
\eqn\corrMuSqSixD{\eqalign{
  {\BB(p)\over \BB_0(p)}\Big|_{\mu^2}= -(\delta^{(1)})^2+i\delta^{(2)}.
}}
The last term $i \delta^{(2)}$ in \corrMuSqSixD\ can be transformed to position space using \posSpaceCont. 
From \PhaseShiftRCB, one finds that the lowest-twist contribution to the second-order phase shift in six dimensions is given by 
\eqn\sixDDouble{
  \delta^{(2),d=6}_0 = {693\pi\over16}{S(4e^{2L}-3)e^{-11L}\over (1-e^{-2L})^3}. 
}
Explicitly, Fourier transforming \sixDDouble, we find the following contribution to the stress tensor sector in the limit $\rho\to\infty$ 

\eqn\sixDCheck{
  \GG^{(2),d=6}(\sigma,\rho)|_{\delta^{(2)}}\rhoequal {i\pi 693p[10,2]\over 4}{e^{-9\rho}\over \sigma}. 
}
This agrees with the imaginary term at $\OO({1\over\sigma})$ after analytically continuing the resummation of minimal-twist double-stress tensors given by Eq.\ (4.8) in \KarlssonDBD.

\listrefs

\bye